\documentclass[pra,twocolumn,showpacs,floatfix]{revtex4}
\usepackage{graphicx}
\begin{document}

\title {Bright solitary waves in a Bose-Einstein condensate and
their interactions}
\author{K. K\"arkk\"ainen$^1$, A. D. Jackson$^2$ and
G. M. Kavoulakis$^3$}
\affiliation{
$^1$Mathematical Physics, Lund Institute of Technology, P.O.
Box 118, SE-22100 Lund, Sweden\\
$^2$Niels Bohr Institute, Blegdamsvej 17, DK-2100,
Copenhagen \O, Denmark \\
$^3$Technological Education Institute of Crete, P.O. Box 1939,
GR-71004, Heraklion, Greece}
\date{\today}

\begin{abstract}

We examine the dynamics of two bright solitary waves with a
negative nonlinear term. The observed repulsion between two
solitary waves -- when these are in an antisymmetric
combination -- is attributed to conservation laws. Slight
breaking of parity, in combination with weak relaxation of
energy, leads the two solitary waves to merge. The effective
repulsion between solitary waves requires certain nearly ideal
conditions and is thus fragile.

\end{abstract}
\maketitle

\section{Introduction}

One of the many interesting features of Bose-Einstein condensed
atoms is that they can support solitary waves, in particular
when these are confined in elongated traps. Under typical
conditions, these gases are very dilute and are described by
the familiar Gross-Pitaevskii equation, a nonlinear
Schr\"odinger equation with an additional term to describe the
external trapping potential.

It is well known that the nonlinear Schr\"odinger equation
(with no external potential) supports solitonic solutions
through the interplay between the nonlinear term and
dispersion. In the presence of an external trapping potential,
the Gross-Pitaevskii equation becomes non-integrable.  In
elongated quasi-one-dimensional traps, it is reasonable to
approximate the three-dimensional solution of the
Gross-Pitaevskii equation by separating longitudinal and
transverse degrees of freedom \cite{JKP}. The resulting
effective one-dimensional nonlinear equation has a nonlinear
term that is not necessarily quadratic \cite{JKP,Sal}. Still,
such nonlinear equations support solitary-wave solutions, which
must be found numerically.

Solitary waves have been created and observed in trapped gases
of atoms \cite{sol1,sol2,sol3,sol4}. In the initial experiments
\cite{sol1,sol2} the effective interaction between the atoms
was repulsive. In this case, the solitary waves are localized
depressions in the density, which are known as ``grey" solitary
waves.  These waves move with a velocity less than the speed of
sound.  When the minimum of the density (at the center of the
wave) becomes zero, they do not move at all and thus become
``dark".

More recently, the two experiments of Refs.\,\cite{sol3,sol4}
considered the case of an effective attraction between the
atoms and observed ``bright" solitary waves, i.e., blob(s) of
atoms which preserve their shape and distinct identity.
Strecker {\it et al.} \cite{sol3} created an initial state of
many separate solitary waves.  While these independent waves
were seen to oscillate in the weak harmonic potential in the
longitudinal direction, they did not merge to form one solitary
wave. In other words, they behaved as if the effective
interaction between two of these waves were repulsive.

Numerous theoretical studies have been motivated by the
experiments of Refs.\,\cite{sol3,sol4}, see e.g.,
Refs.\,\cite{Stoof,Carr,Sal2}. Reference \cite{Stoof} offered
an explanation for the observed effective repulsion between
solitary waves. As argued there, the experiments had been
performed in a manner that gave rise to a phase difference of
$\pi$ between adjacent solitary waves. According to an older
study \cite{Gordon}, solitary waves with a phase difference
equal to $\pi$ indeed repel each other.

In the present study we use a toroidal trap \cite{CCR} as a
model for examining the time evolution of a system that
initially has two solitary waves using numerical solutions to
the corresponding time-dependent one-dimensional
Gross-Pitaevskii equation. Remarkably, such toroidal traps have
been designed \cite{Kurn}, and very recently persistent
currents have been created and observed in such traps
\cite{phil}. The basic conclusion of our study is that the
effective repulsion between solitary waves is due to
conservation laws and thus fragile.

In what follows, we first present our model in Sec II. In
Sec.\,III we examine the dynamics of the gas in the case of
weak dissipation, starting with perfectly
symmetric/antisymmetric initial conditions and with no external
potential along the torus. We observe that the symmetric
configuration of two blobs merges on a short time scale; the
blobs in the initially antisymmetric configuration remain
distinct and separated.  Using these results as ``reference"
plots, we examine in Sec.\,IV the effect of a weak random
potential on perfectly symmetric/antisymmetric initial
conditions. We also examine in Sec.\,V the time evolution of
states that deviate slightly from perfect symmetry/antisymmetry
in the absence of any random potential. In both cases, the
symmetric (or nearly symmetric) initial configuration shows
essentially the same behavior as the reference symmetric
system. On the other hand, the antisymmetric configuration with
the addition of an extra weak random potential and the nearly
antisymmetric configuration with no external potential both
lead to a merger of the two blobs after a moderate transient
time. In the antisymmetric case, the final state is strongly
influenced by weak deviations from the ``ideal" case. Finally,
in Sec.VI we discuss our results.

\begin{figure}[t]
\includegraphics[width=5.5cm,height=5.5cm]{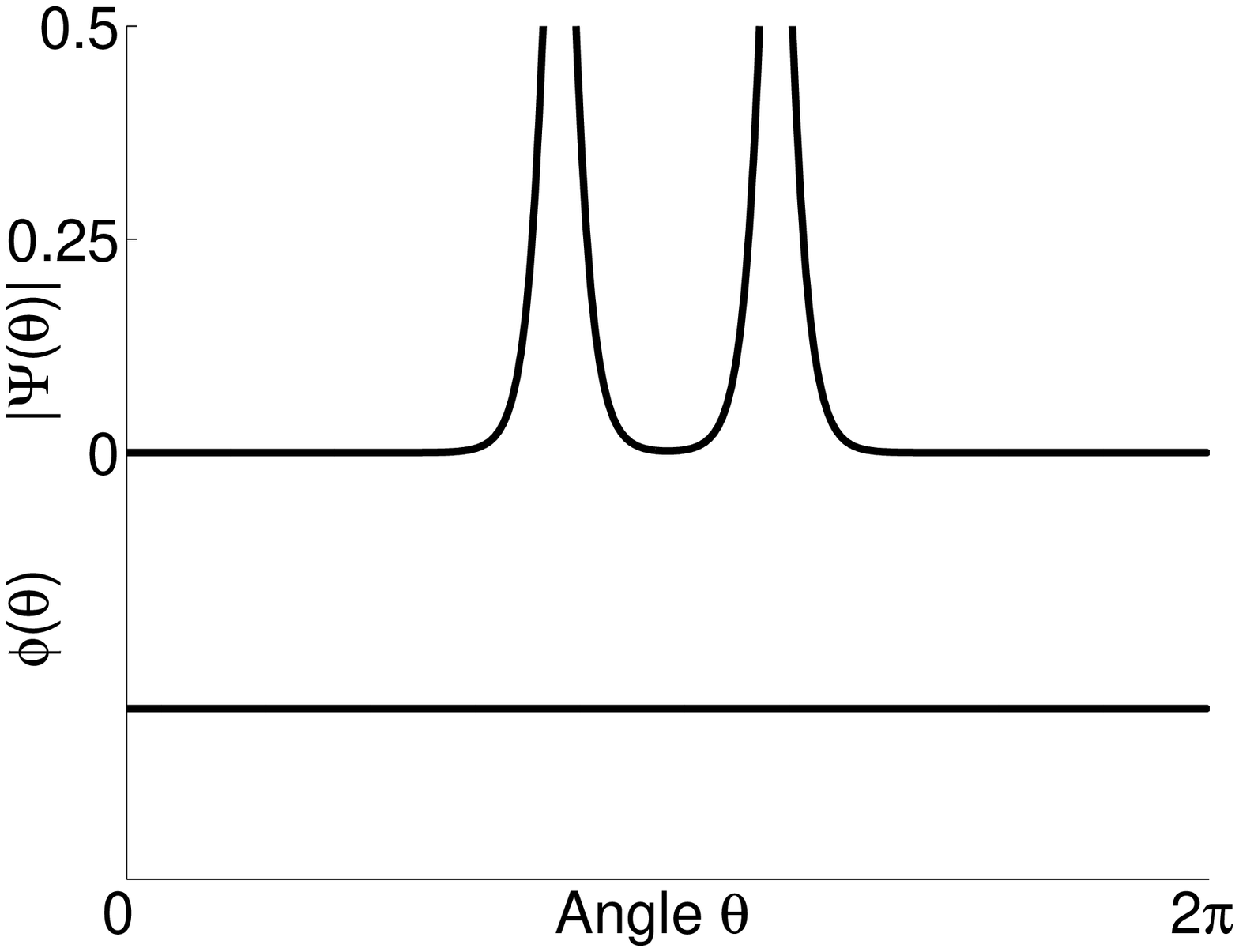}
\includegraphics[width=5.5cm,height=5.5cm]{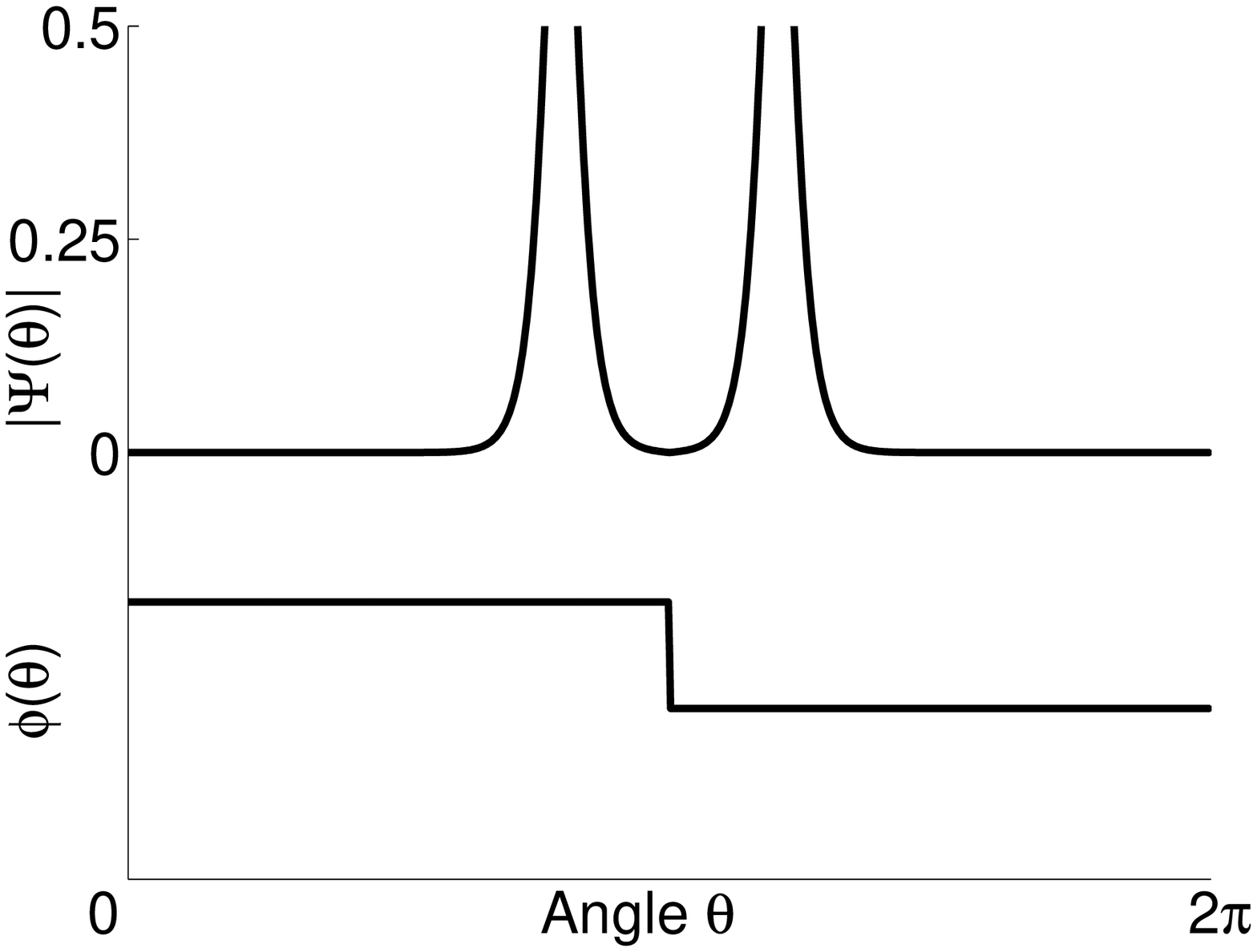}
\caption[]{Plot of $|\Psi(\theta, t=0)|$ and of $\phi(\theta, t
= 0)$ for the symmetric (higher) and for the antisymmetric
(lower) case, $\alpha = \pm 1$ in Eq.\,(\ref{initialcond}).}
\label{FIG1}
\end{figure}

\section{Model}

We consider a tight toroidal trap and use the mean-field
approximation. Tight confinement along the cross section of the
torus allows us to assume that the transverse degrees of
freedom are frozen, and thus the corresponding time-dependent
order parameter $\Psi(\theta, t)$ satisfies the
(one-dimensional) equation
\begin{eqnarray}
   i \hbar \frac {\partial \Psi} {\partial t}
   = \frac {\hbar^2} {2M R^2}
[-\frac {\partial^2 \Psi} {\partial \theta^2}
+ g |\Psi|^2 \Psi + V(\theta) \Psi],
\end{eqnarray}
where $g = 8 \pi N a R/S$, and $V(\theta)$ is the external
potential measured in units of $E_0 = \hbar^2/(2 M R^2)$. Here,
$M$ is the atomic mass, $R$ is the radius of the torus, $N$ is
the atom number, $a$ is the scattering length (which is taken
to be negative), and $S$ is the cross section of the torus. The
total length of the torus is chosen to be $16 \pi$ in our
simulations.

\begin{figure}[t]
\includegraphics[width=2.5cm,height=2.5cm]{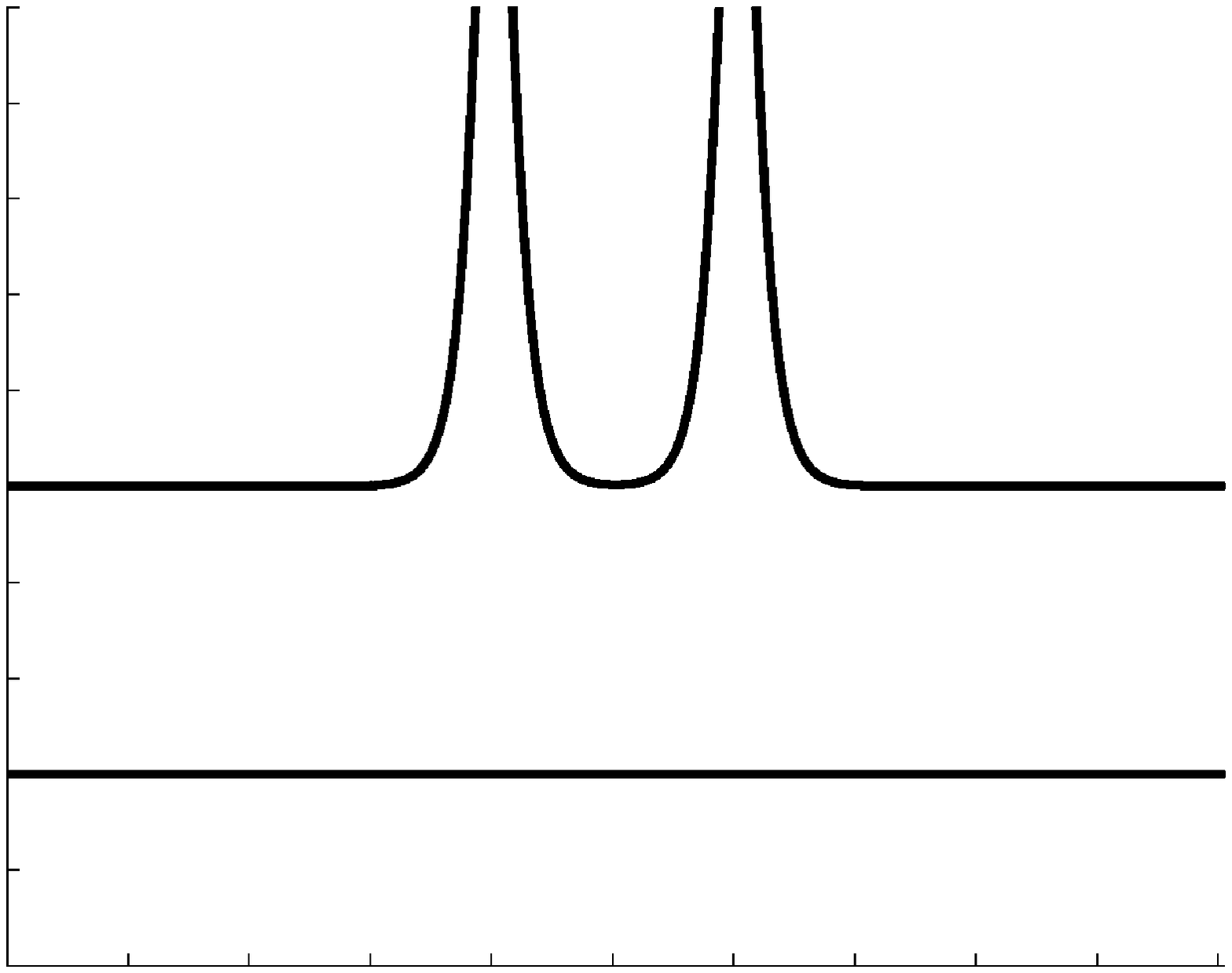}
\includegraphics[width=2.5cm,height=2.5cm]{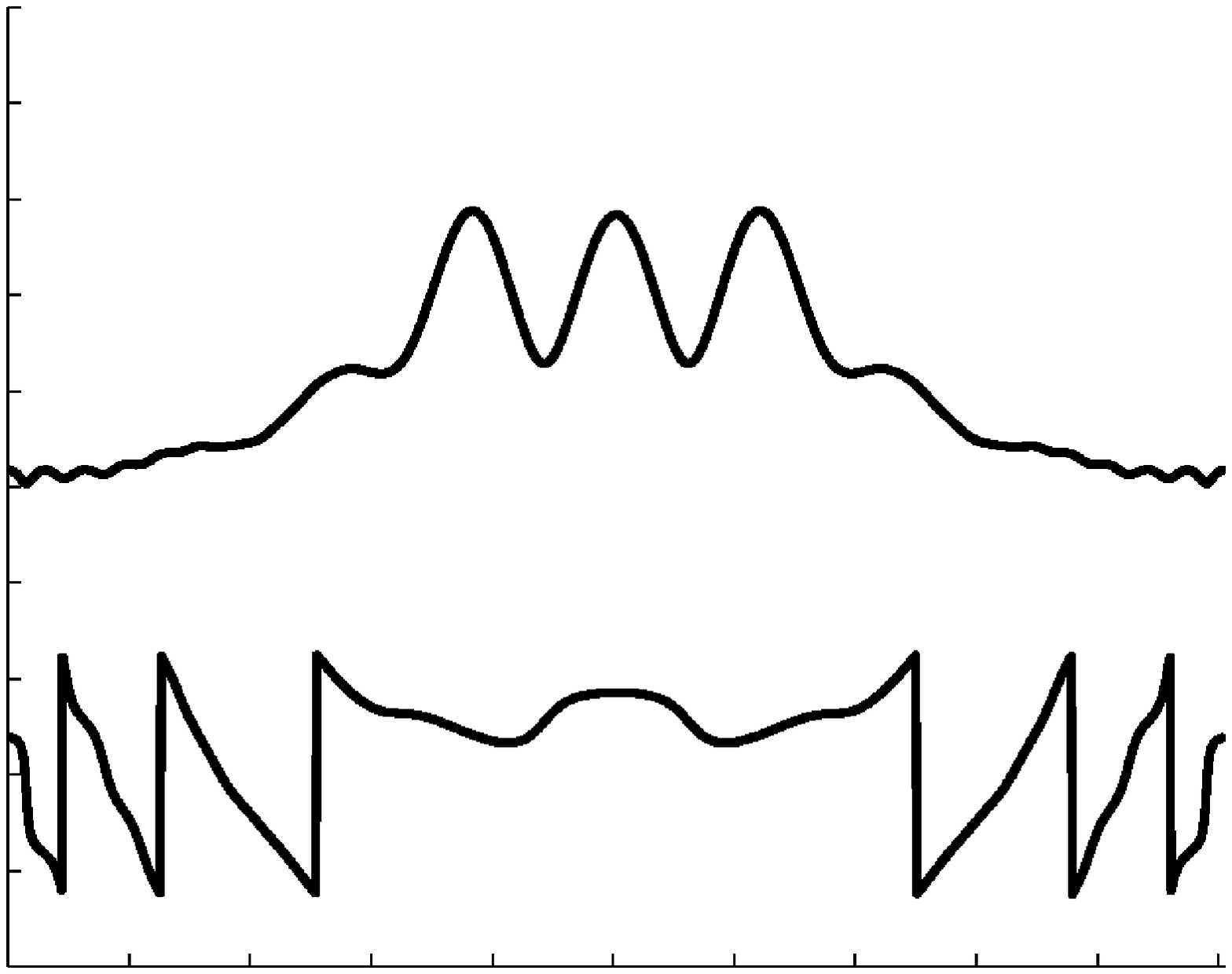}
\includegraphics[width=2.5cm,height=2.5cm]{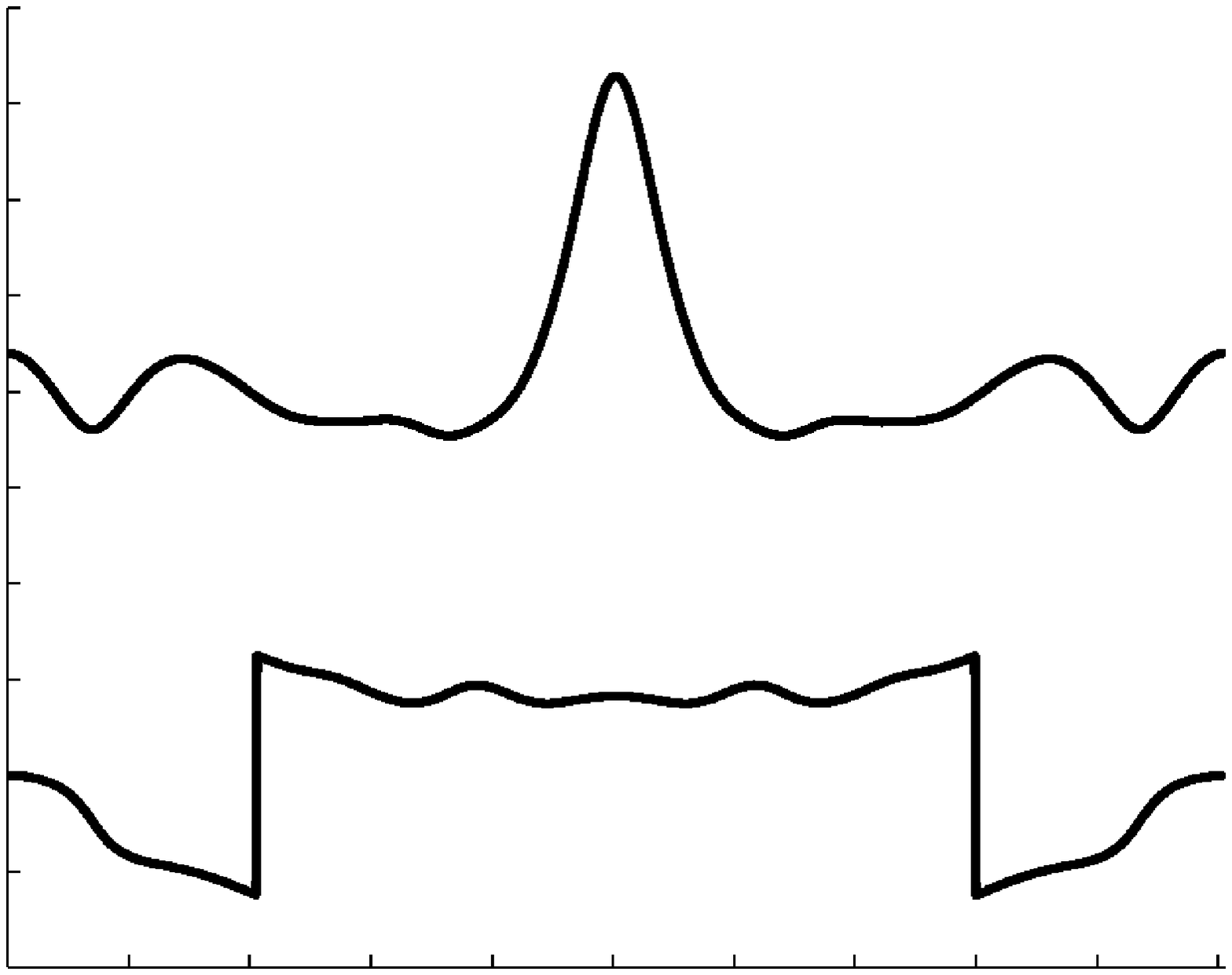}
\includegraphics[width=2.5cm,height=2.5cm]{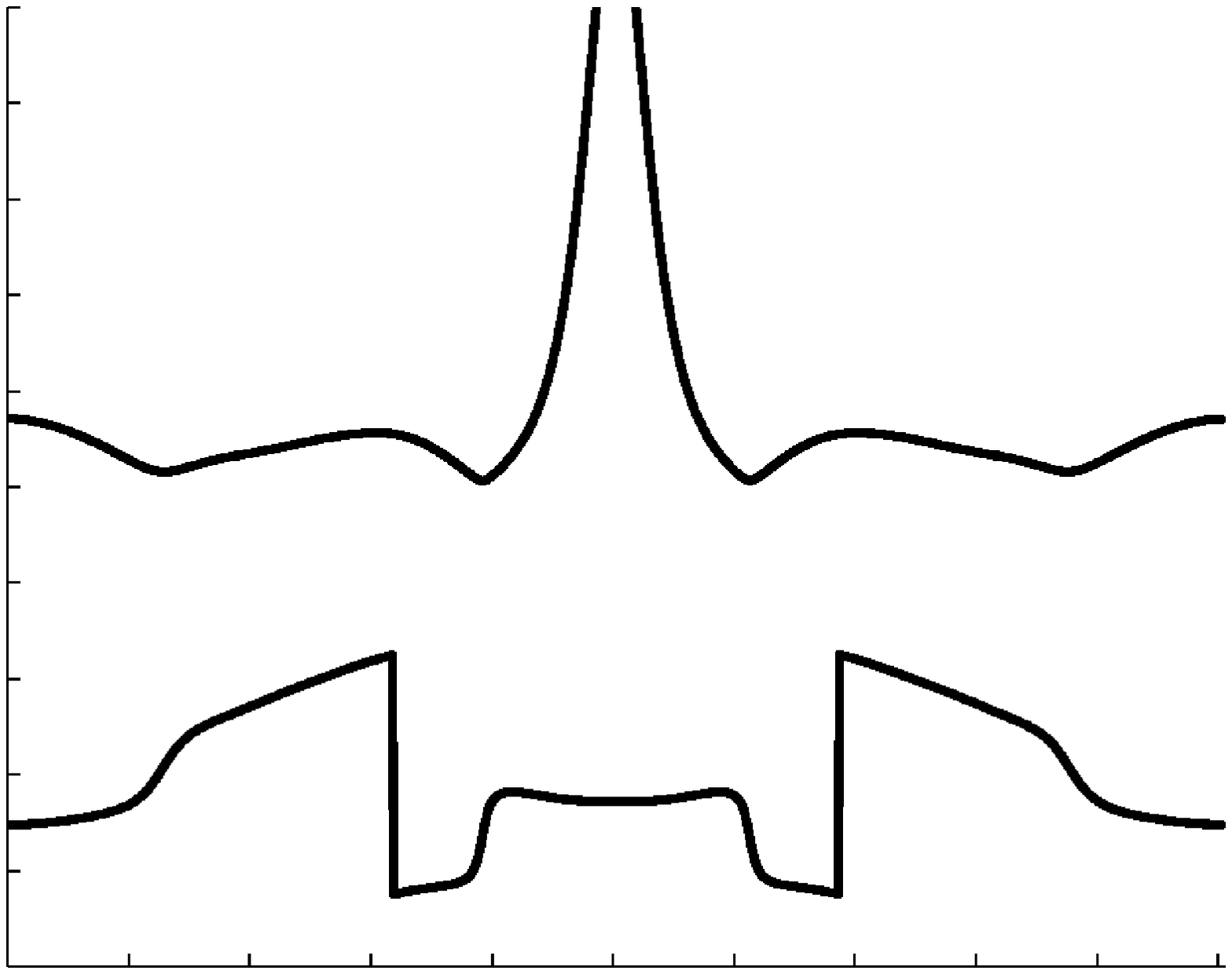}
\includegraphics[width=2.5cm,height=2.5cm]{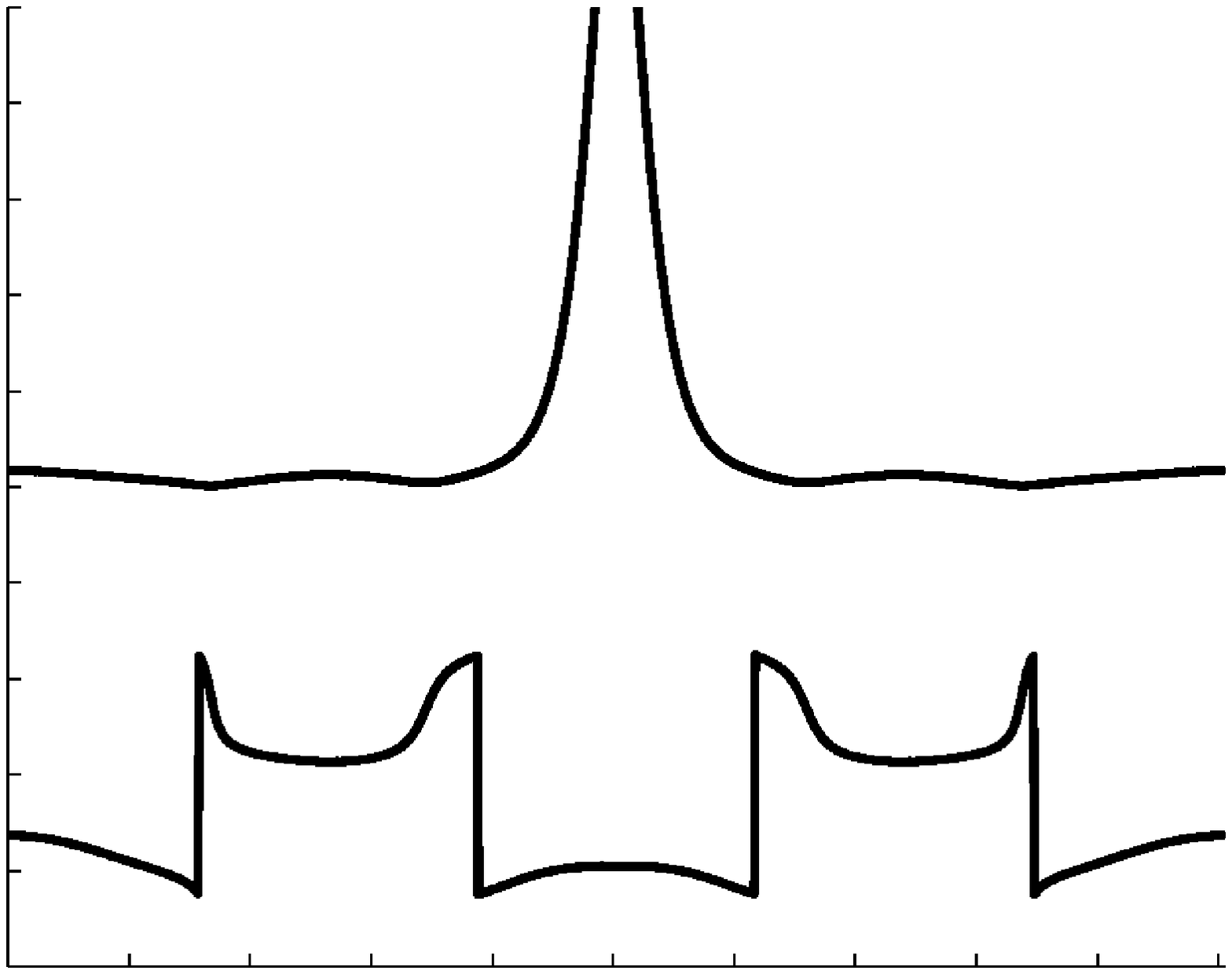}
\caption[]{Snapshots of $|\Psi(\theta, t)|$ (higher curve in
each panel) and of $\phi(\theta, t)$ (lower curve in each
panel) of the order parameter $\Psi = |\Psi(\theta,t)| e^{i
\phi(\theta,t)}$, for the symmetric initial configuration,
$\alpha = 1$, for $t/t_0=0, 10, 50, 100$, and $150$. The
axes are the same as in Fig.\,1. In all the above graphs
there is no external potential, $V = 0$.}
\label{FIG2}
\end{figure}

\begin{figure}[t]
\includegraphics[width=5.5cm,height=4.5cm]{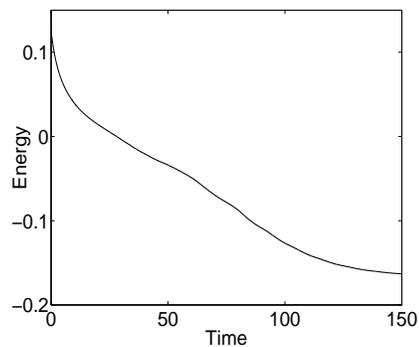}
\caption[]{The energy of the system in units of $E_0 =
\hbar^2/(2MR^2)$, versus time for $0 \le t/t_0 \le 150$
corresponding to Fig.\,\ref{FIG2}.}
\label{FIG3}
\end{figure}

As shown in Refs.\,\cite{Ueda,GMK}, below a critical (negative)
value of the parameter $g$, there is an instability from a
state of homogeneous density to a state with localized density
that breaks the rotational invariance of the Hamiltonian. This
localized state corresponds to a solitary wave, and the
critical value of $g$ is $g_c = -\pi$ for the parameters chosen
here. We adopt the value of $g = -4 < g_c$ in all simulations
below. Therefore, the lowest energy state is a single solitary
wave.

We add an extra term on the left side of the above equation
to model dissipation and write
\begin{eqnarray}
   (i - \gamma) \hbar \frac {\partial \Psi} {\partial t}
   = \frac {\hbar^2} {2M R^2}
[-\frac {\partial^2 \Psi} {\partial \theta^2}
+ g |\Psi|^2 \Psi + V(\theta) \Psi].
\end{eqnarray}
The real positive dimensionless parameter $\gamma$ describes
the ``strength" of dissipation. Since we solve an initial value
problem, we also need to specify the initial condition. This is
\begin{equation}
   \Psi(\theta, t=0) = \frac 1 {\sqrt{1 + \alpha^2}}
   [\psi(\theta - \theta_0) +
   \alpha \psi(\theta + \theta_0)].
\label{initialcond}
\end{equation}
Here $\psi(\theta) = \lambda/\cosh(\lambda \theta)$, with
$\lambda = 3/2$, is a static, well localized blob. We choose
$\theta_0 = 2 \pi/5$ so that the two blobs are reasonably
distinct but still have a small overlap, as shown in the graphs
of Fig.\,\ref{FIG1}, for $\alpha = \pm 1$.

\begin{figure}[t]
\includegraphics[width=2.5cm,height=2.5cm]{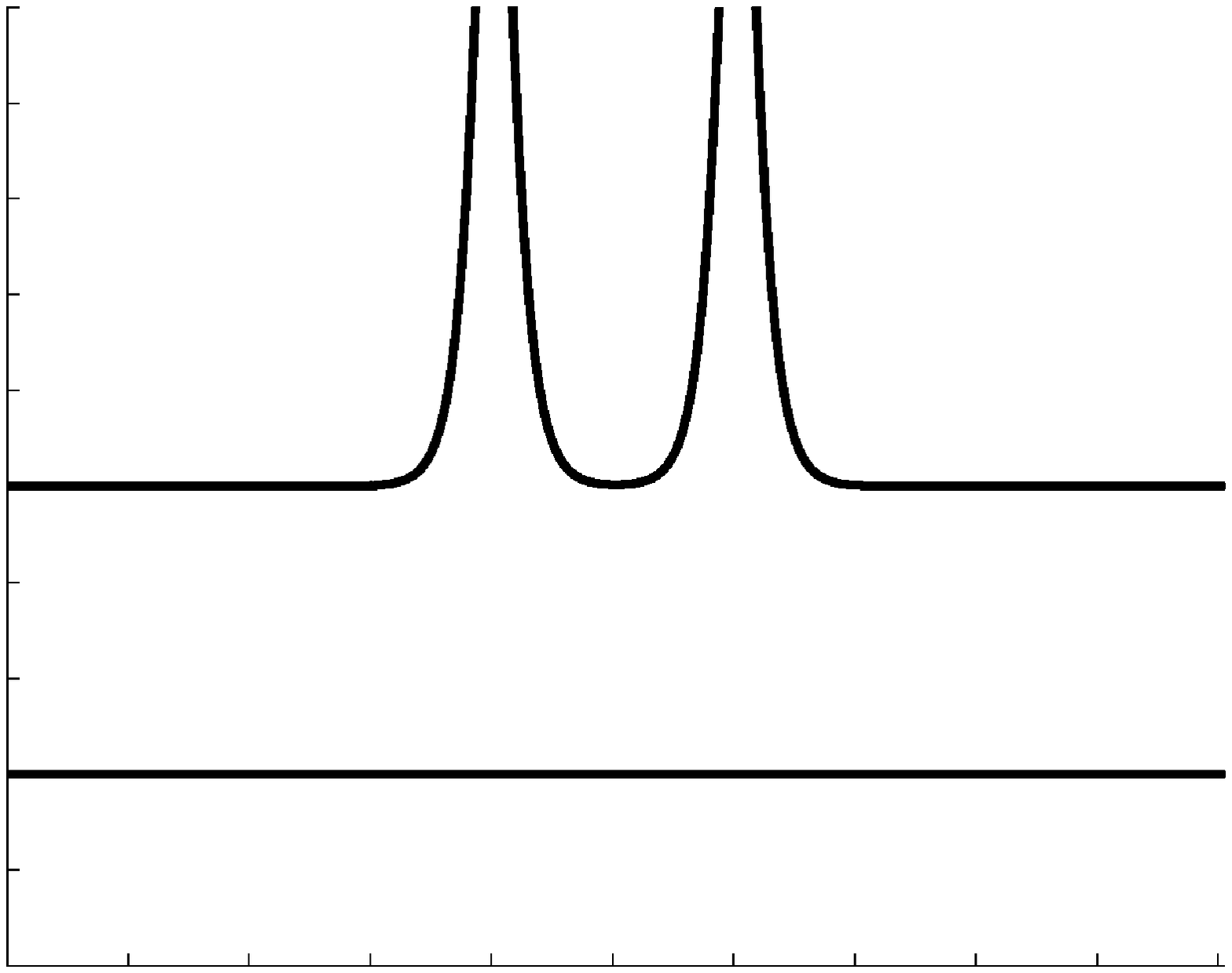}
\includegraphics[width=2.5cm,height=2.5cm]{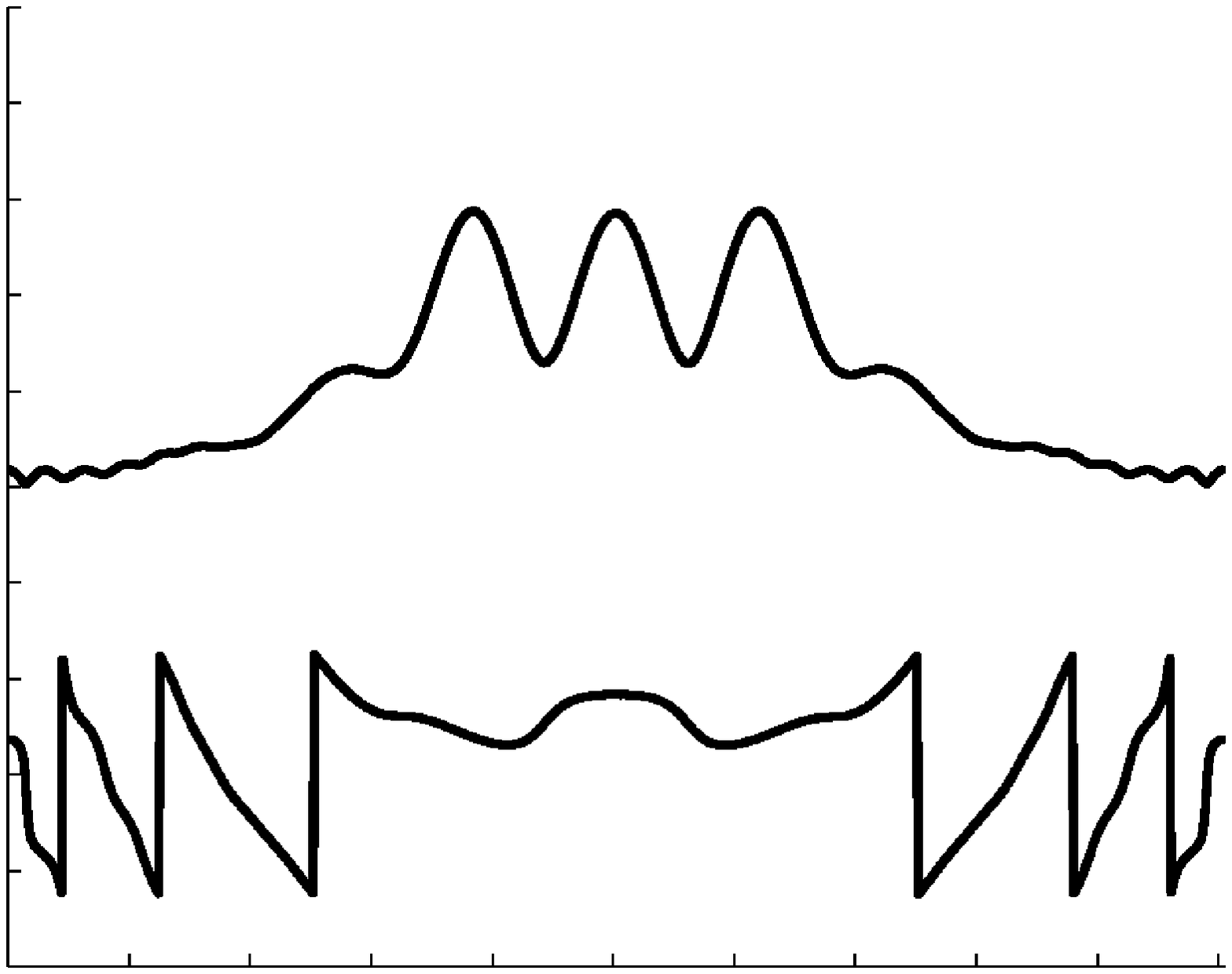}
\includegraphics[width=2.5cm,height=2.5cm]{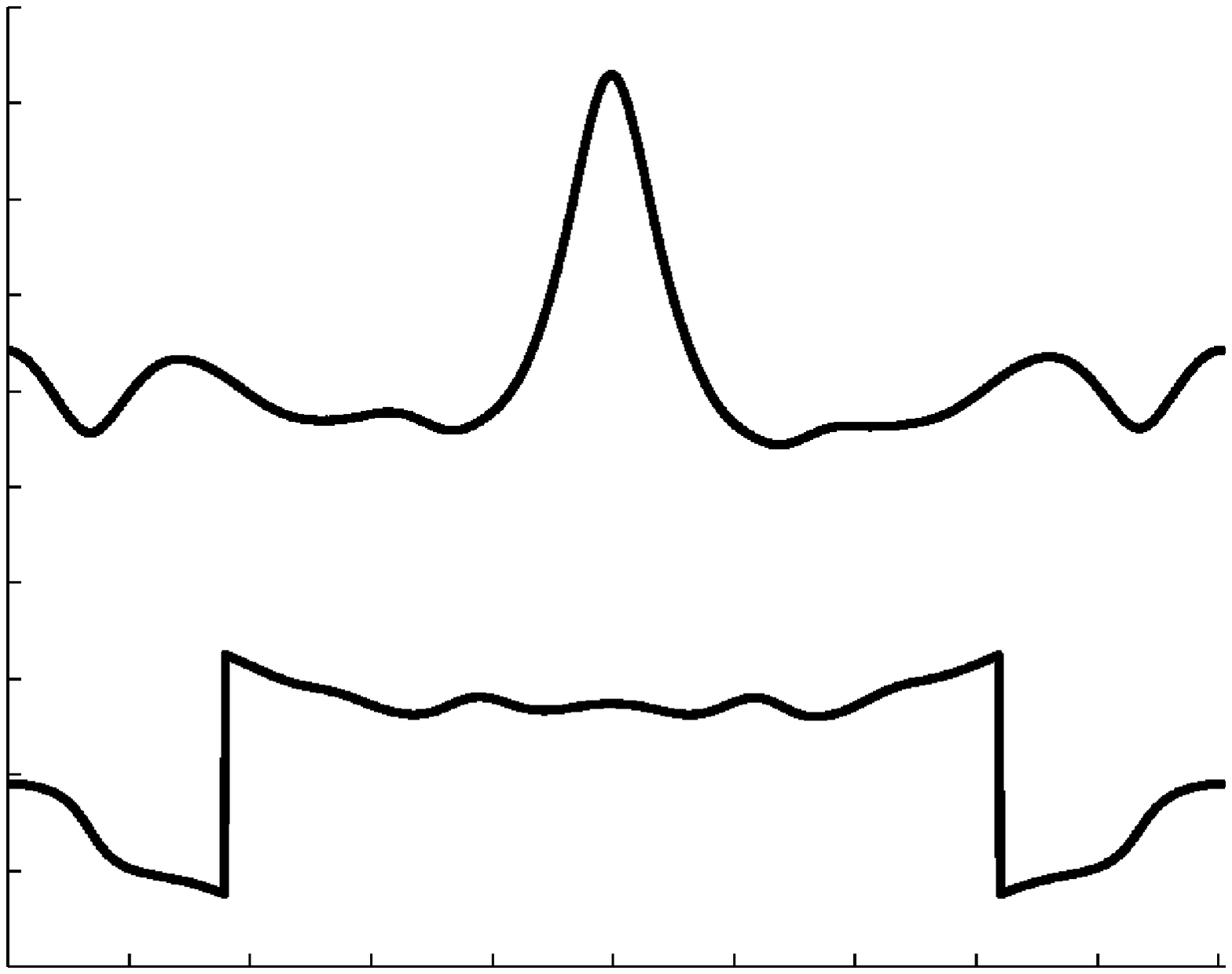}
\includegraphics[width=2.5cm,height=2.5cm]{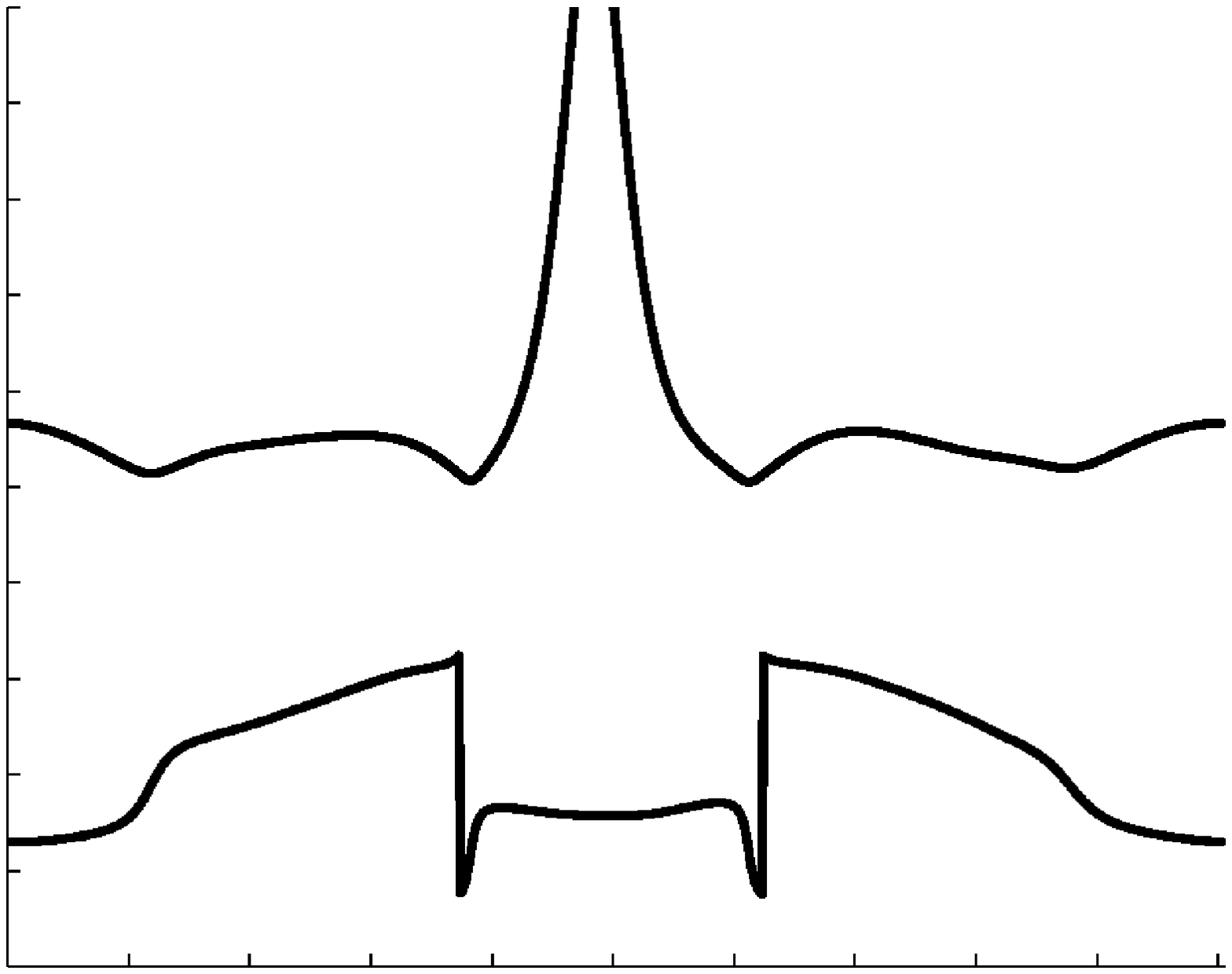}
\includegraphics[width=2.5cm,height=2.5cm]{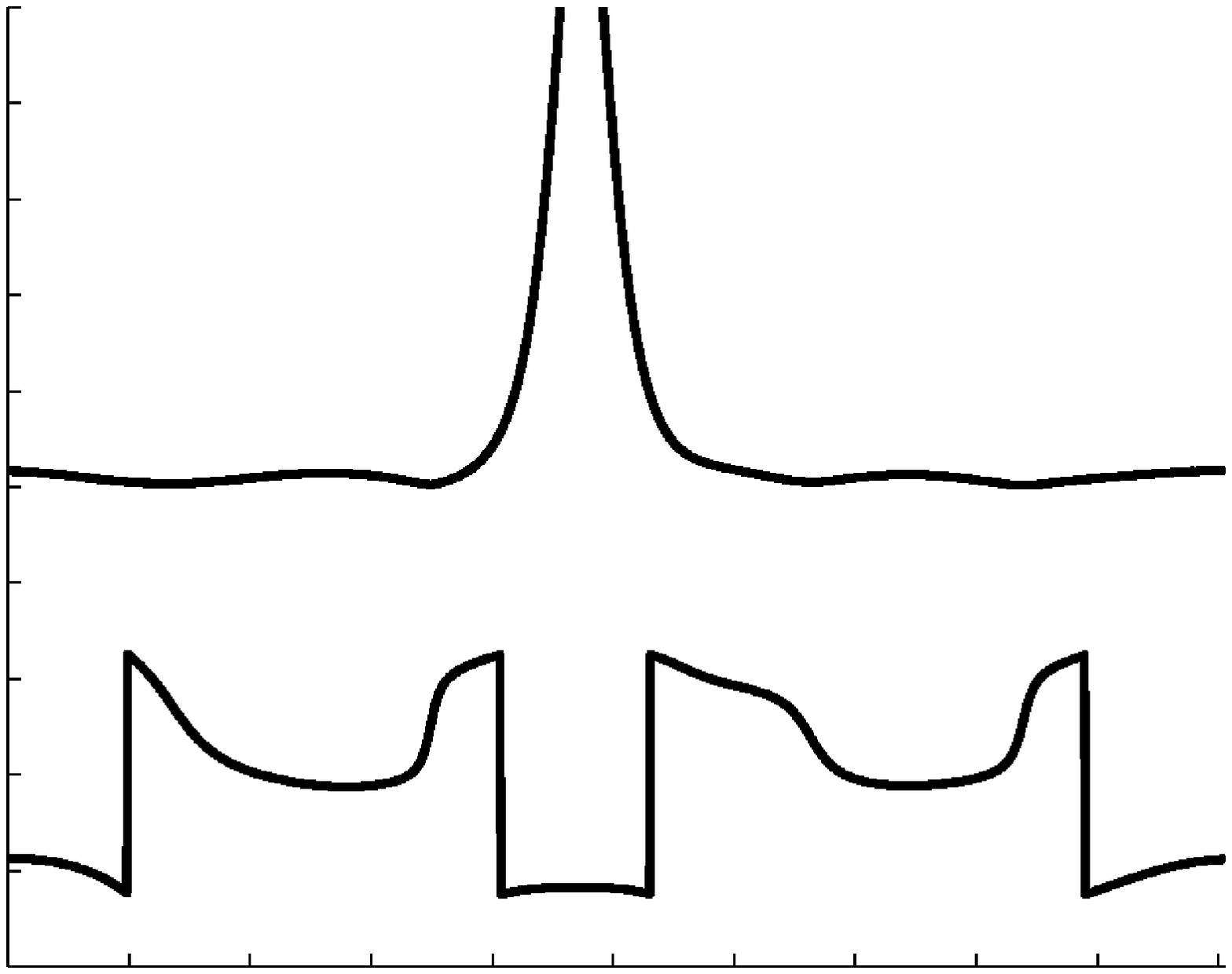}
\caption[]{Snapshots of $|\Psi(\theta, t)|$ (higher curve in
each panel) and of $\phi(\theta, t)$ (lower curve in each
panel) of the order parameter $\Psi =|\Psi(\theta,t)| e^{i
\phi(\theta,t)}$, for the symmetric initial condition, for
$t/t_0=0, 10, 50, 100$, and $150$, for a weak random potential
(shown in Fig.\,\ref{FIG14}), with a symmetric initial
configuration, $\alpha = 1$. The axes are the same as in
Fig.\,1.}
\label{FIG4}
\end{figure}

\begin{figure}[t]
\includegraphics[width=5.5cm,height=4.5cm]{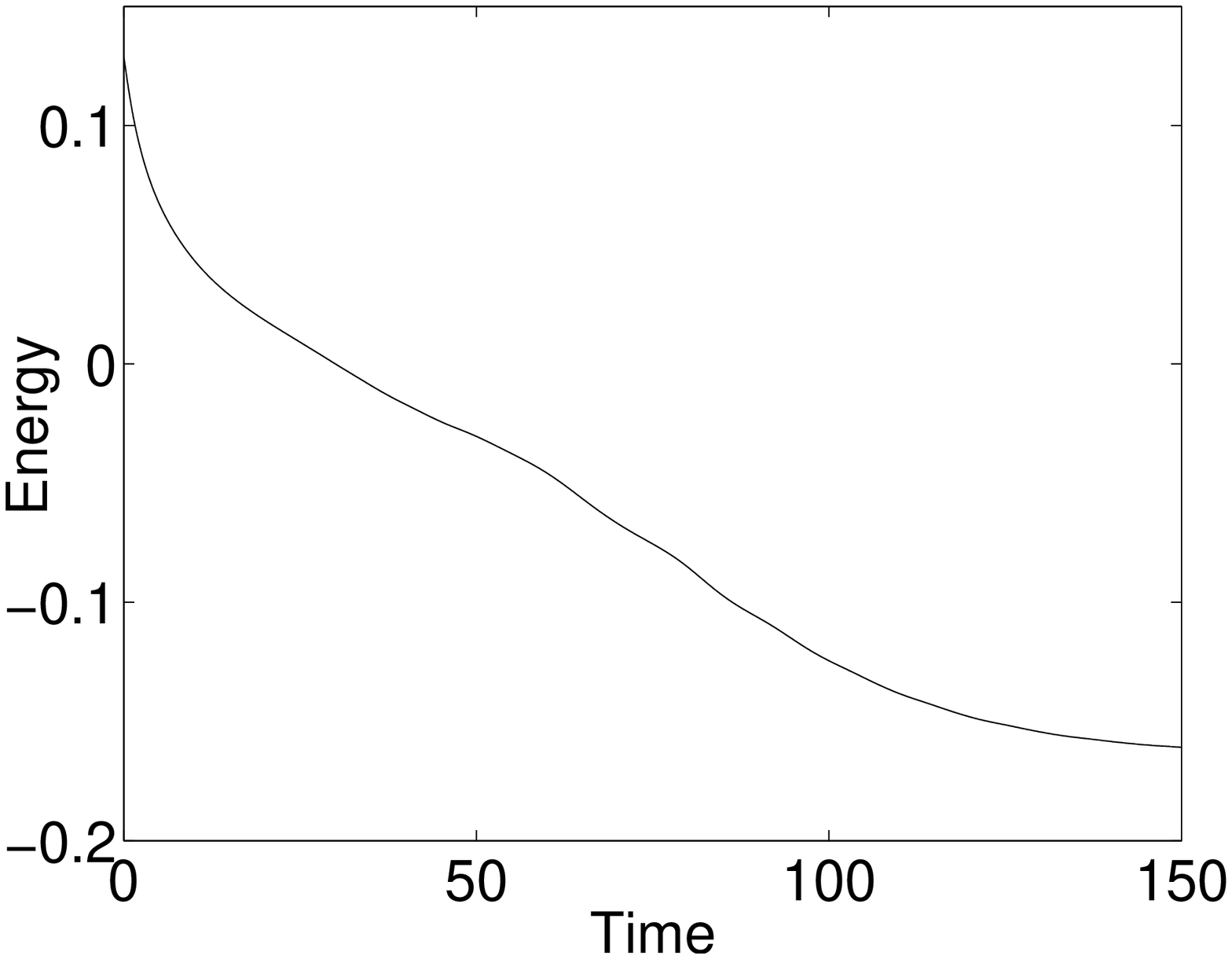}
\caption[]{The energy of the system in units of $E_0 =
\hbar^2/(2MR^2)$, versus time for $0 \le t/t_0 \le 150$
corresponding to Fig.\,\ref{FIG4}.}
\label{FIG5}
\end{figure}

\begin{figure}[t]
\includegraphics[width=2.5cm,height=2.5cm]{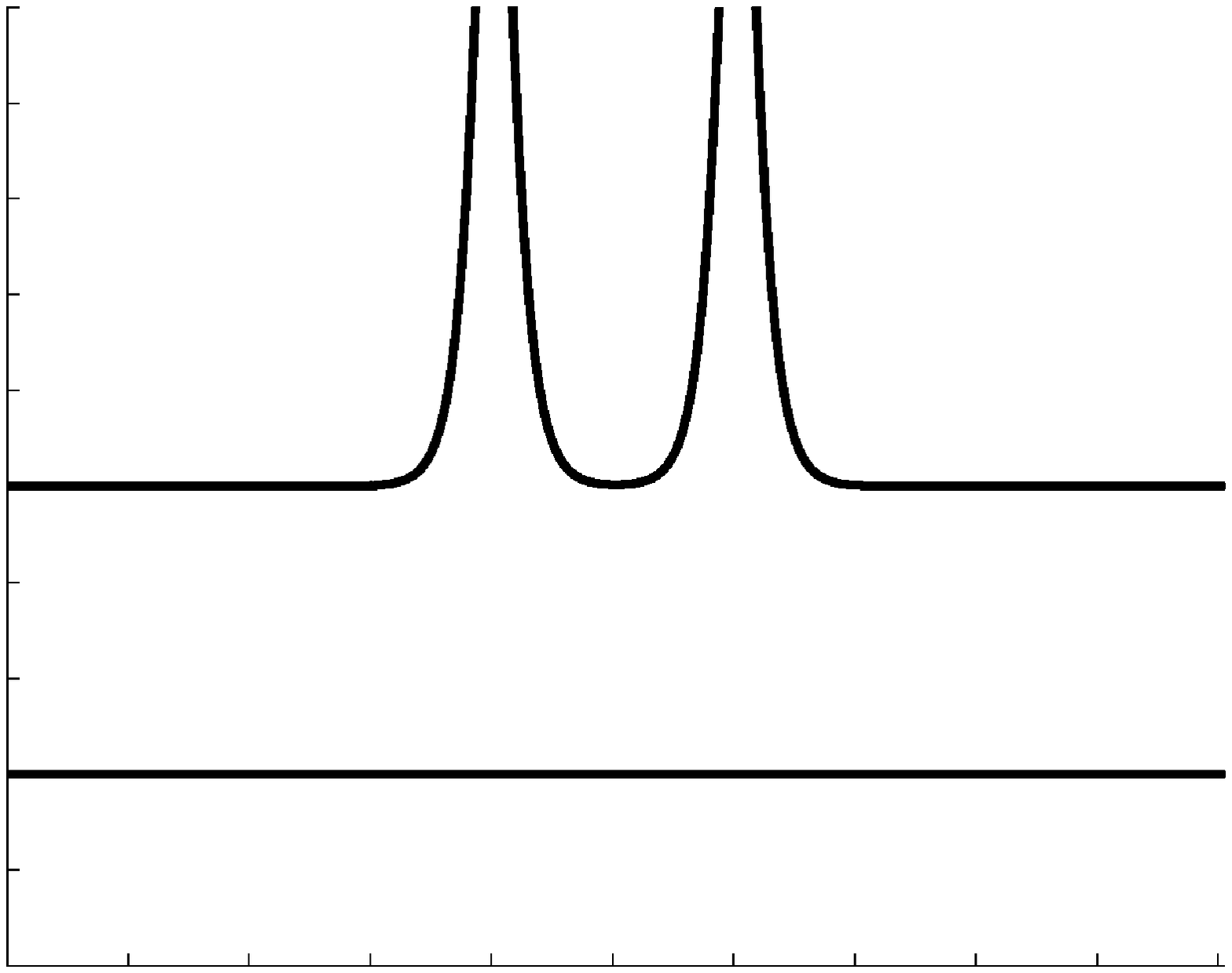}
\includegraphics[width=2.5cm,height=2.5cm]{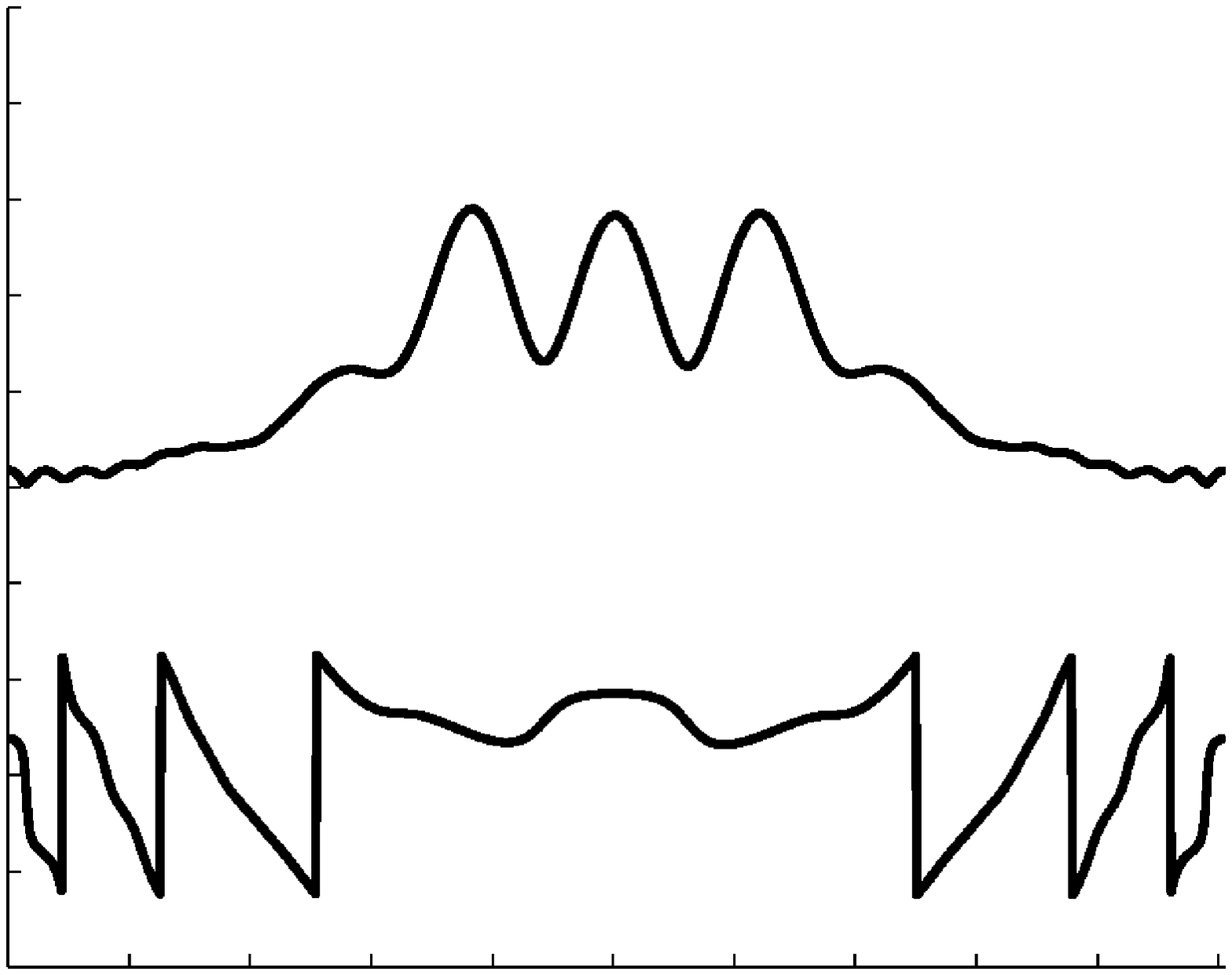}
\includegraphics[width=2.5cm,height=2.5cm]{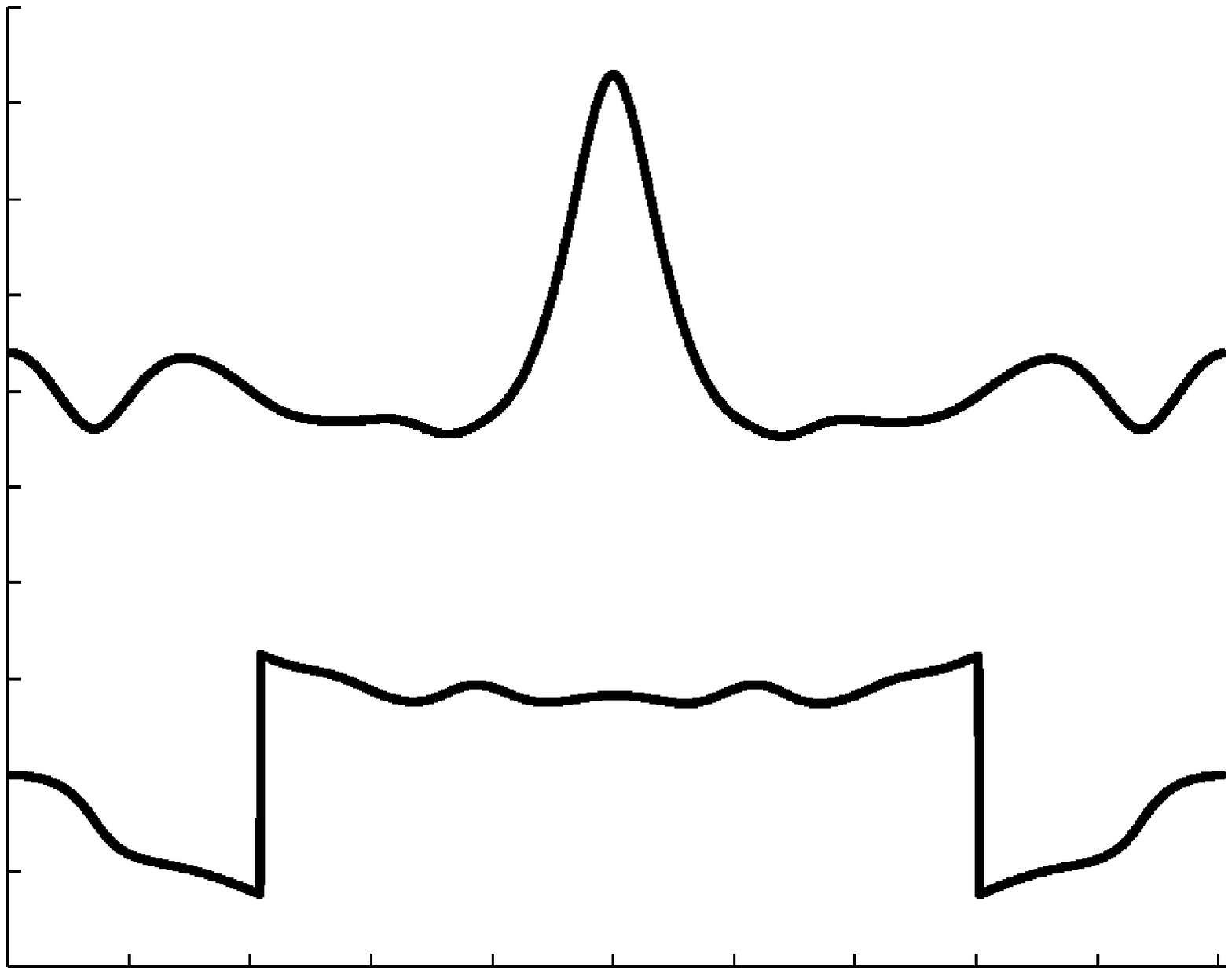}
\includegraphics[width=2.5cm,height=2.5cm]{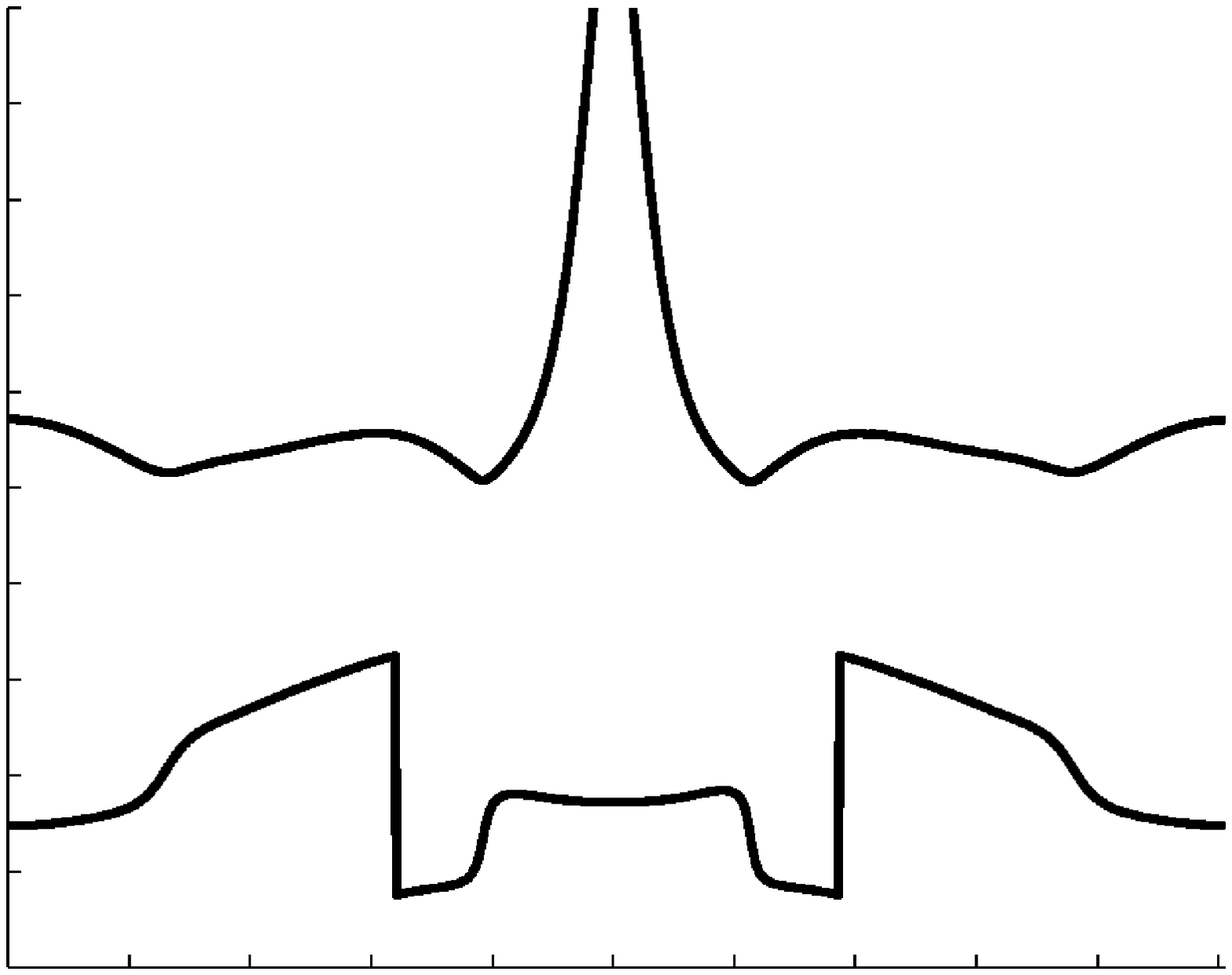}
\includegraphics[width=2.5cm,height=2.5cm]{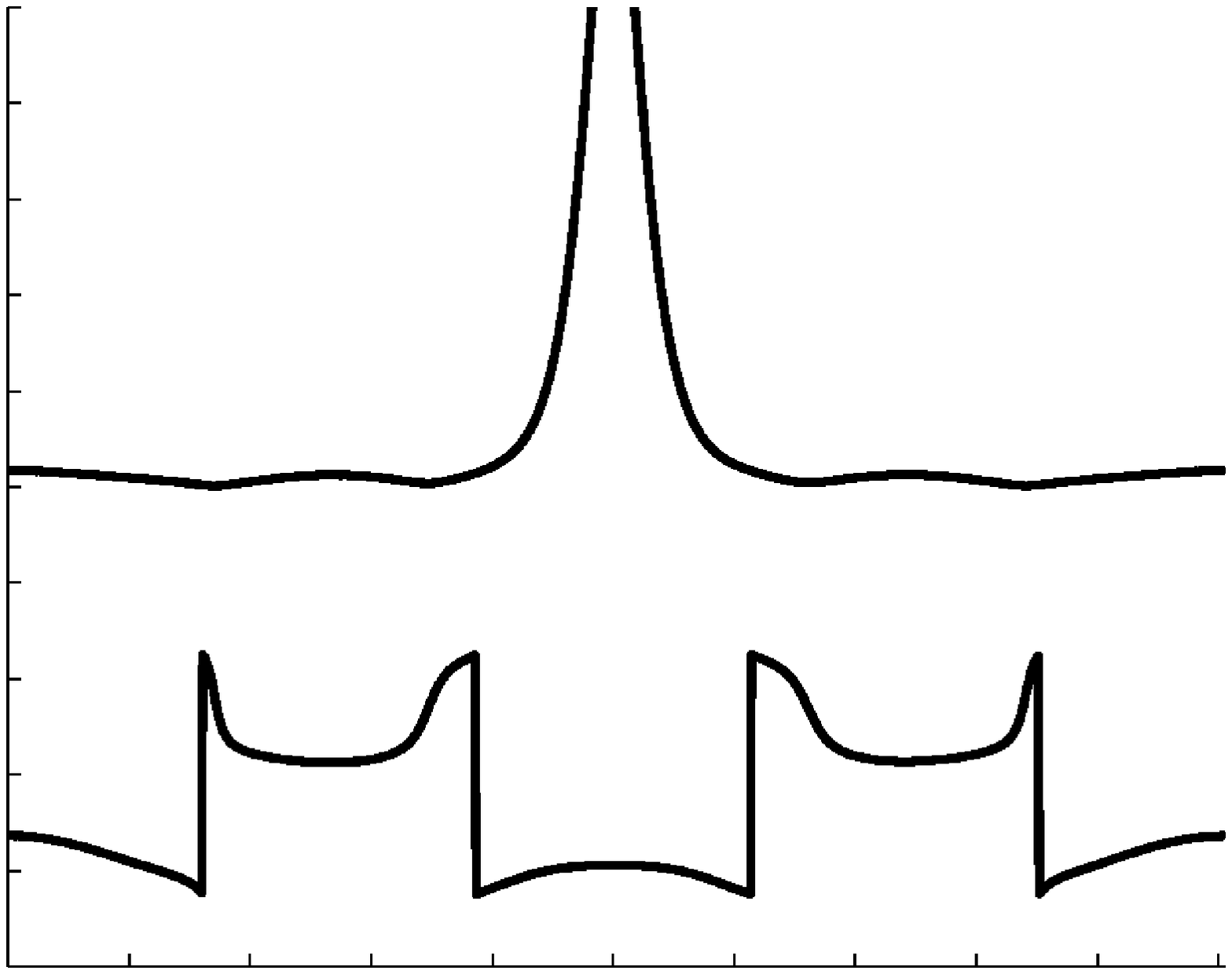}
\caption[]{Snapshots of $|\Psi(\theta, t)|$ (higher curve in
each panel) and of $\phi(\theta, t)$ (lower curve in each
panel) of the order parameter $\Psi =|\Psi(\theta,t)| e^{i
\phi(\theta,t)}$, for the almost symmetric initial condition,
$\alpha = 1.01$, for $t/t_0=0, 10, 50, 100$, and $150$. The
axes are the same as in Fig.\,1. In all the above graphs there
is no external potential, $V = 0$.}
\label{FIG6}
\end{figure}

\begin{figure}[t]
\includegraphics[width=5.5cm,height=4.5cm]{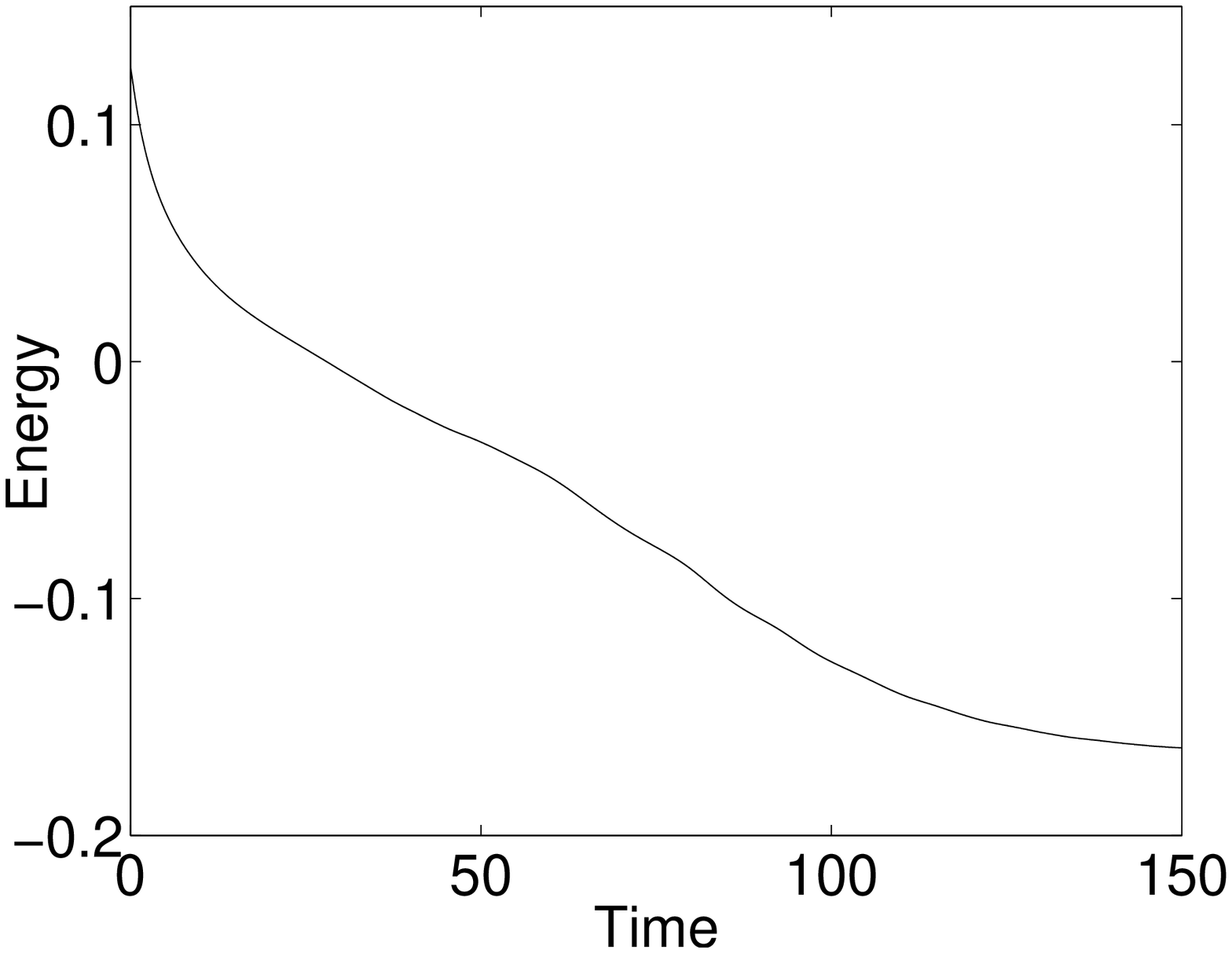}
\caption[]{The energy of the system in units of $E_0 =
\hbar^2/(2MR^2)$, versus time for $0 \le t/t_0 \le 150$
corresponding to Fig.\,\ref{FIG6}.}
\label{FIG7}
\end{figure}

\begin{figure}[t]
\includegraphics[width=2.5cm,height=2.5cm]{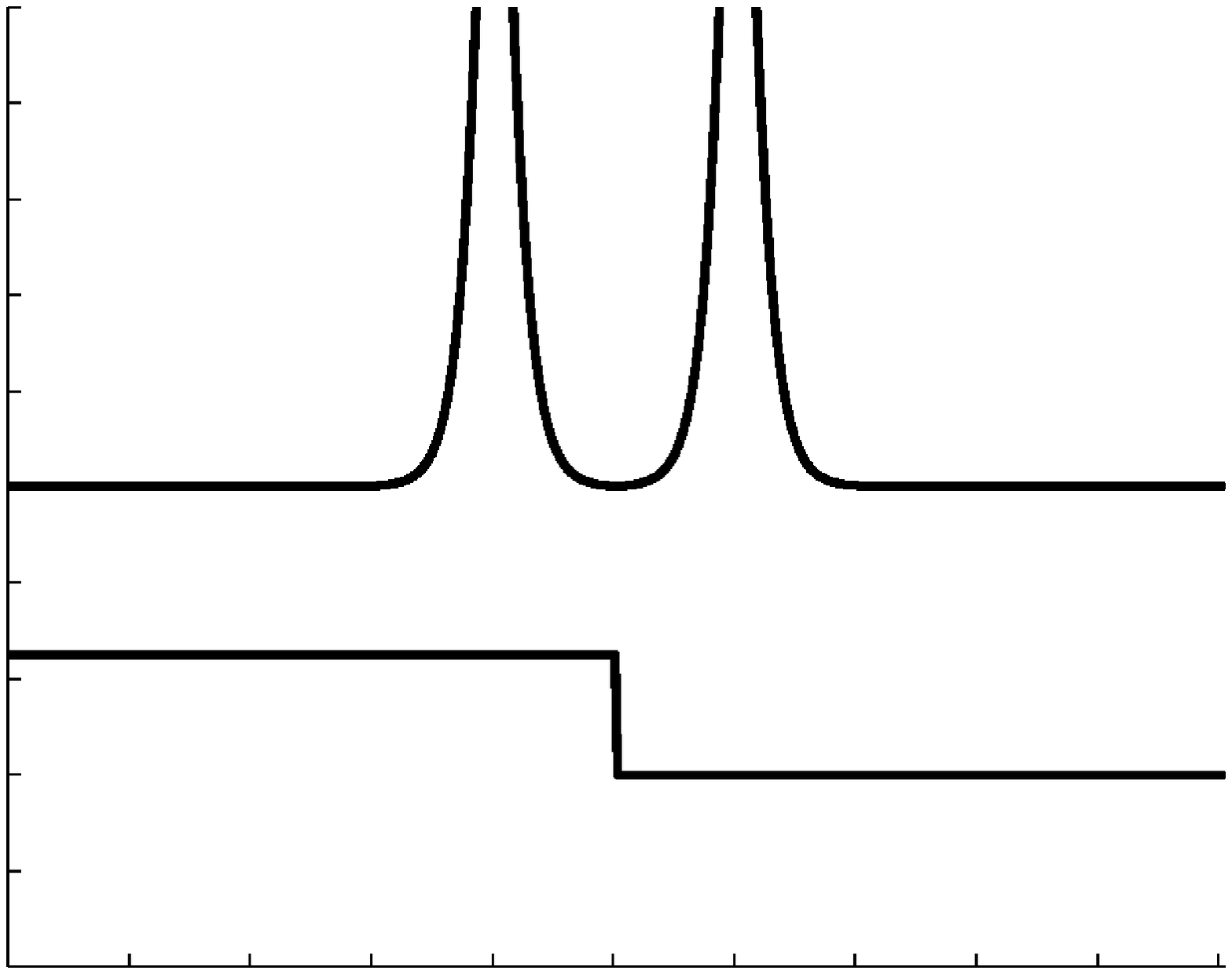}
\includegraphics[width=2.5cm,height=2.5cm]{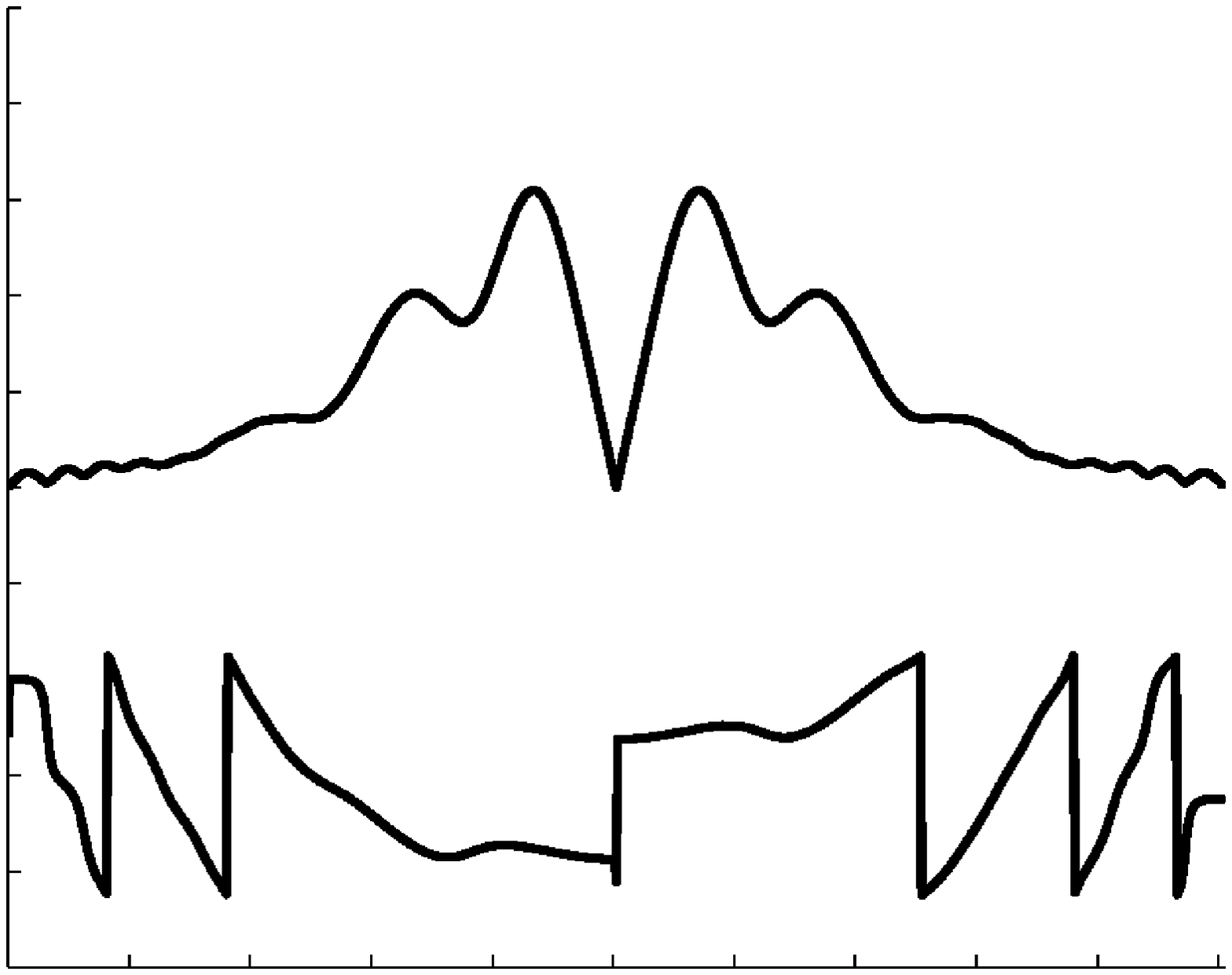}
\includegraphics[width=2.5cm,height=2.5cm]{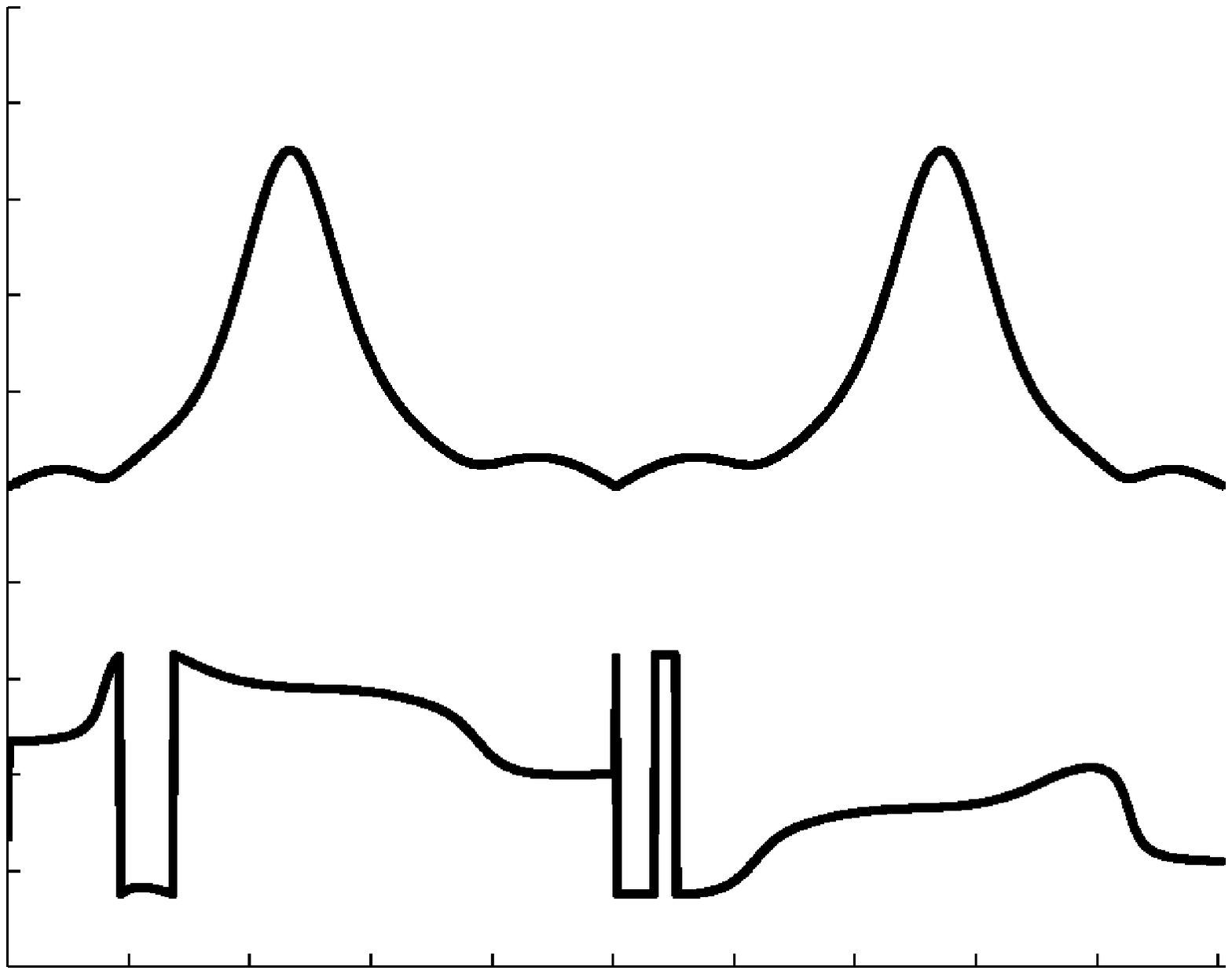}
\includegraphics[width=2.5cm,height=2.5cm]{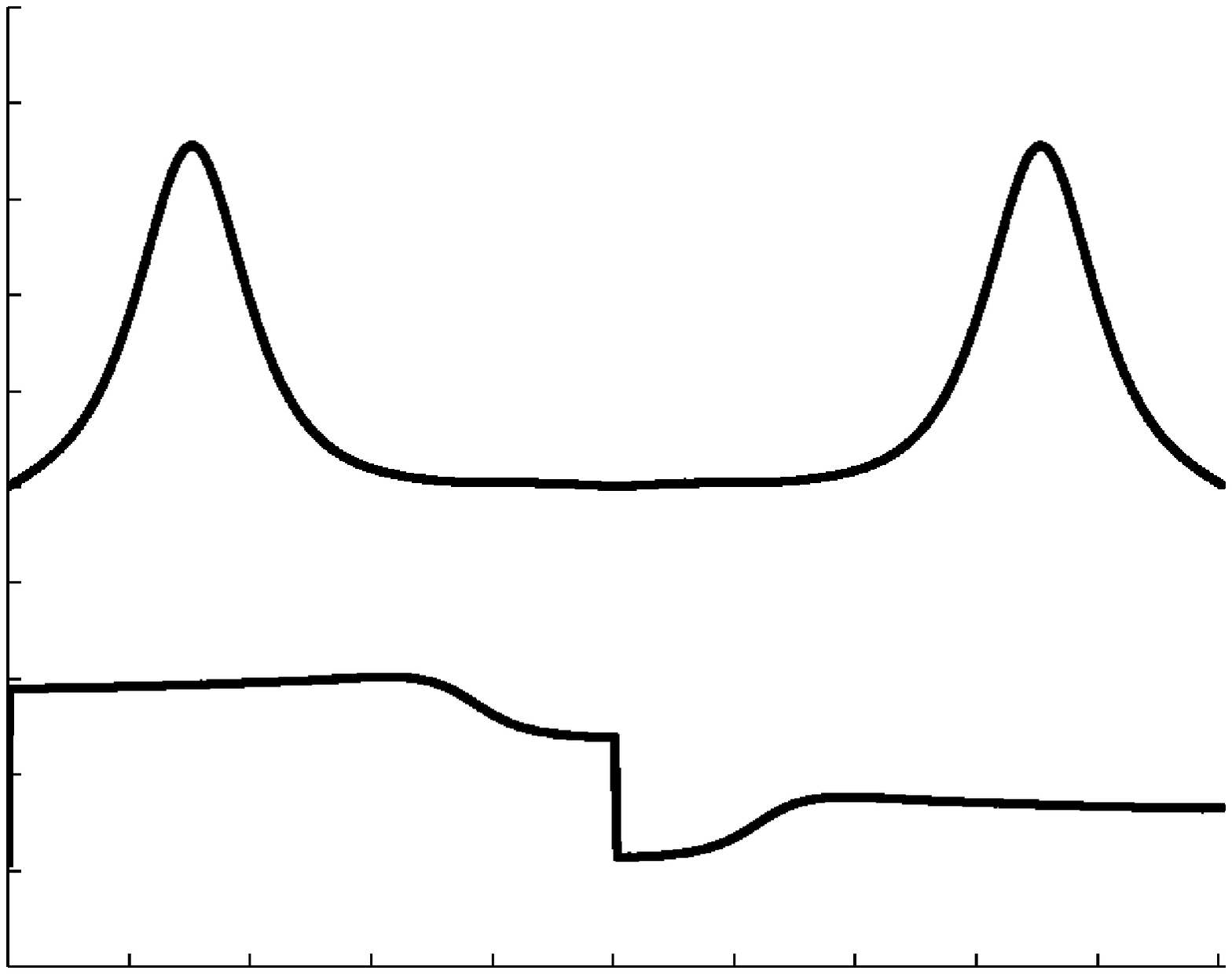}
\includegraphics[width=2.5cm,height=2.5cm]{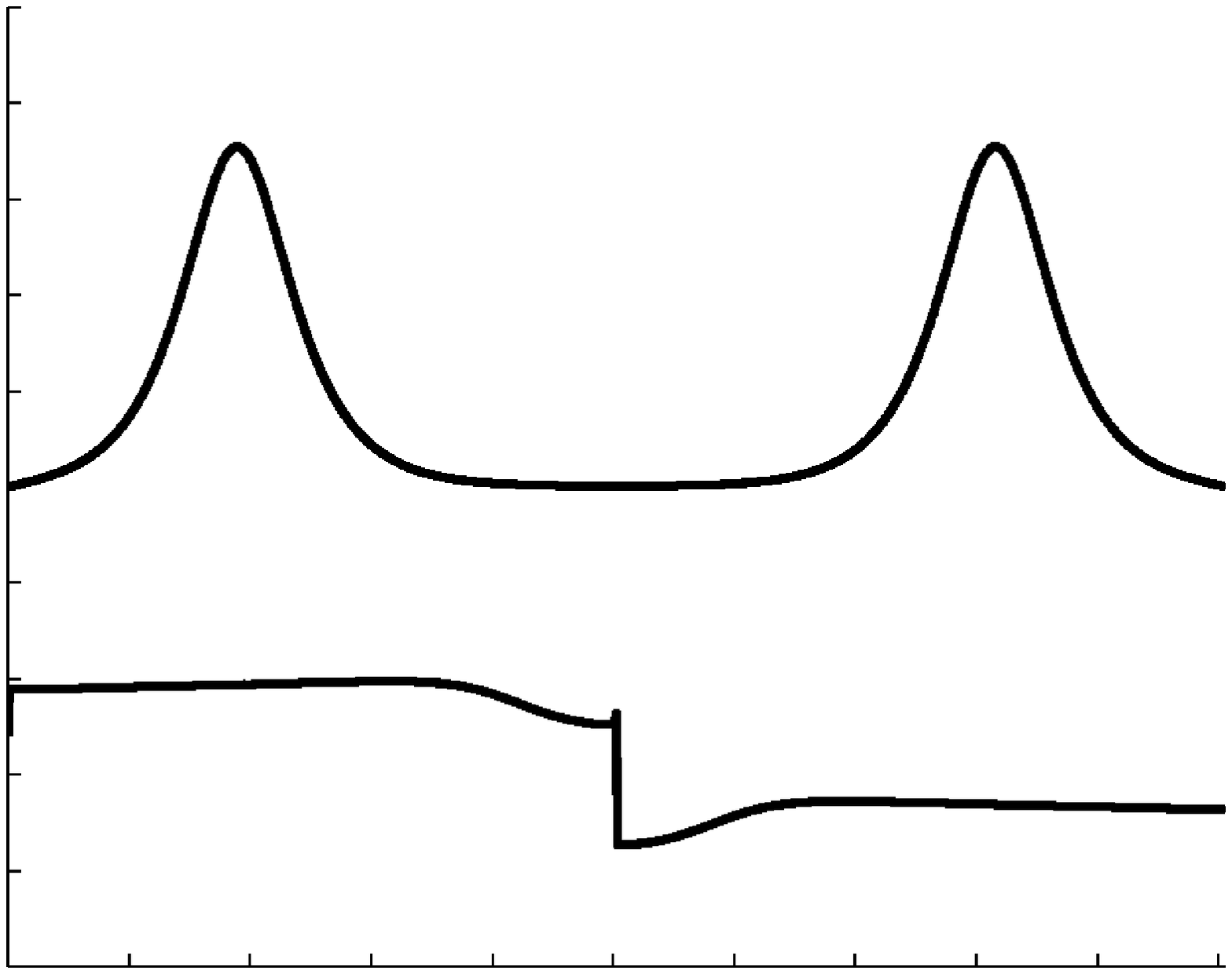}
\caption[]{Snapshots of $|\Psi(\theta, t)|$ (higher curve in
each panel) and of $\phi(\theta, t)$ (lower curve in each
panel) of the order parameter $\Psi =|\Psi(\theta,t)| e^{i
\phi(\theta,t)}$, for the antisymmetric initial configuration,
$\alpha = -1$, for $t/t_0=0, 10, 100, 300$, and $400$. The axes
are the same as in Fig.\,1. In all the above graphs there is no
external potential, $V = 0$.}
\label{FIG8}
\end{figure}

\begin{figure}[t]
\includegraphics[width=5.5cm,height=4.5cm]{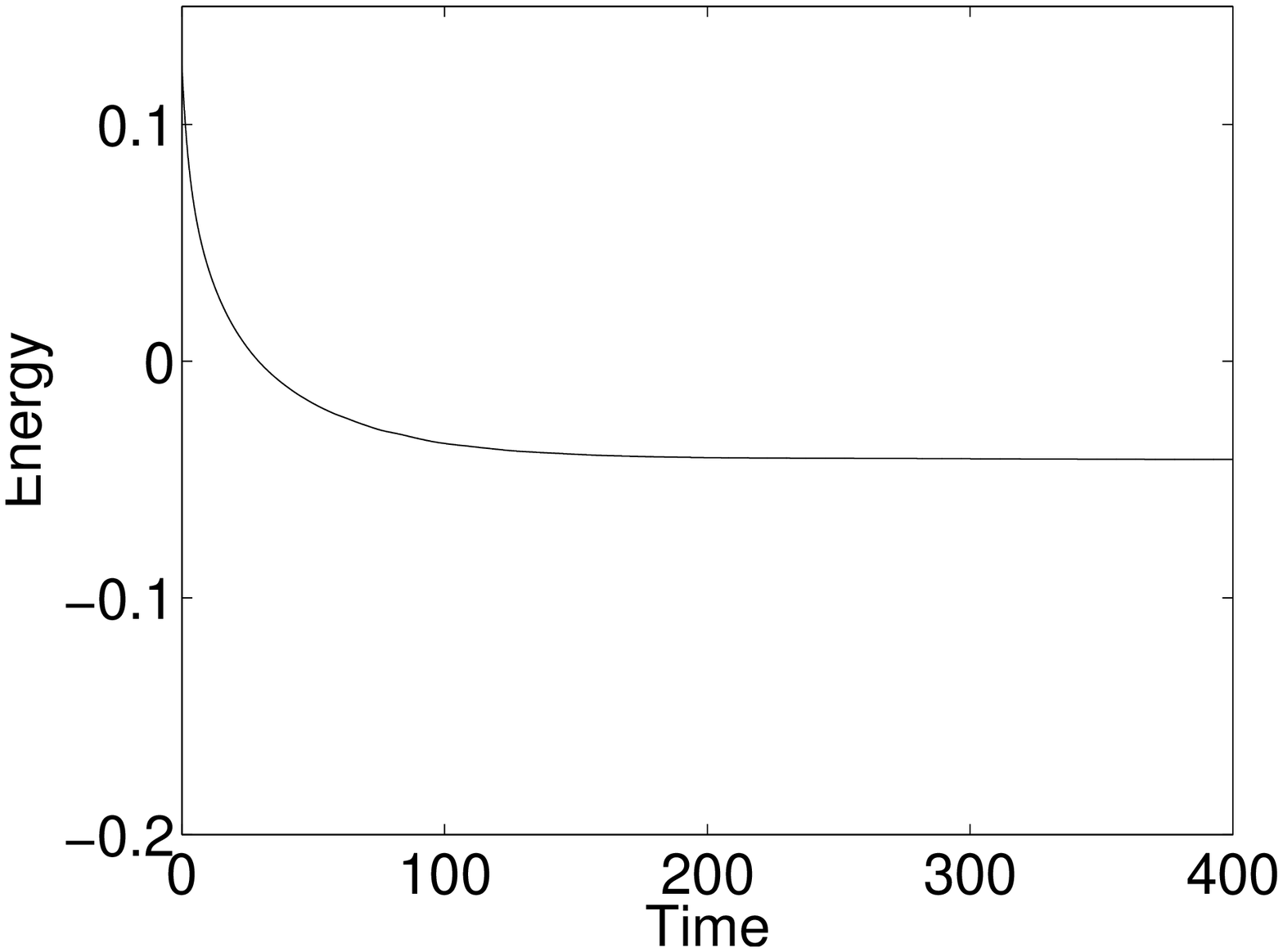}
\caption[]{The energy of the system in units of $E_0 =
\hbar^2/(2MR^2)$, versus time for $0 \le t/t_0 \le 400$
corresponding to Fig.\,\ref{FIG8}.}
\label{FIG9}
\end{figure}

\begin{figure}[t]
\includegraphics[width=2.5cm,height=2.5cm]{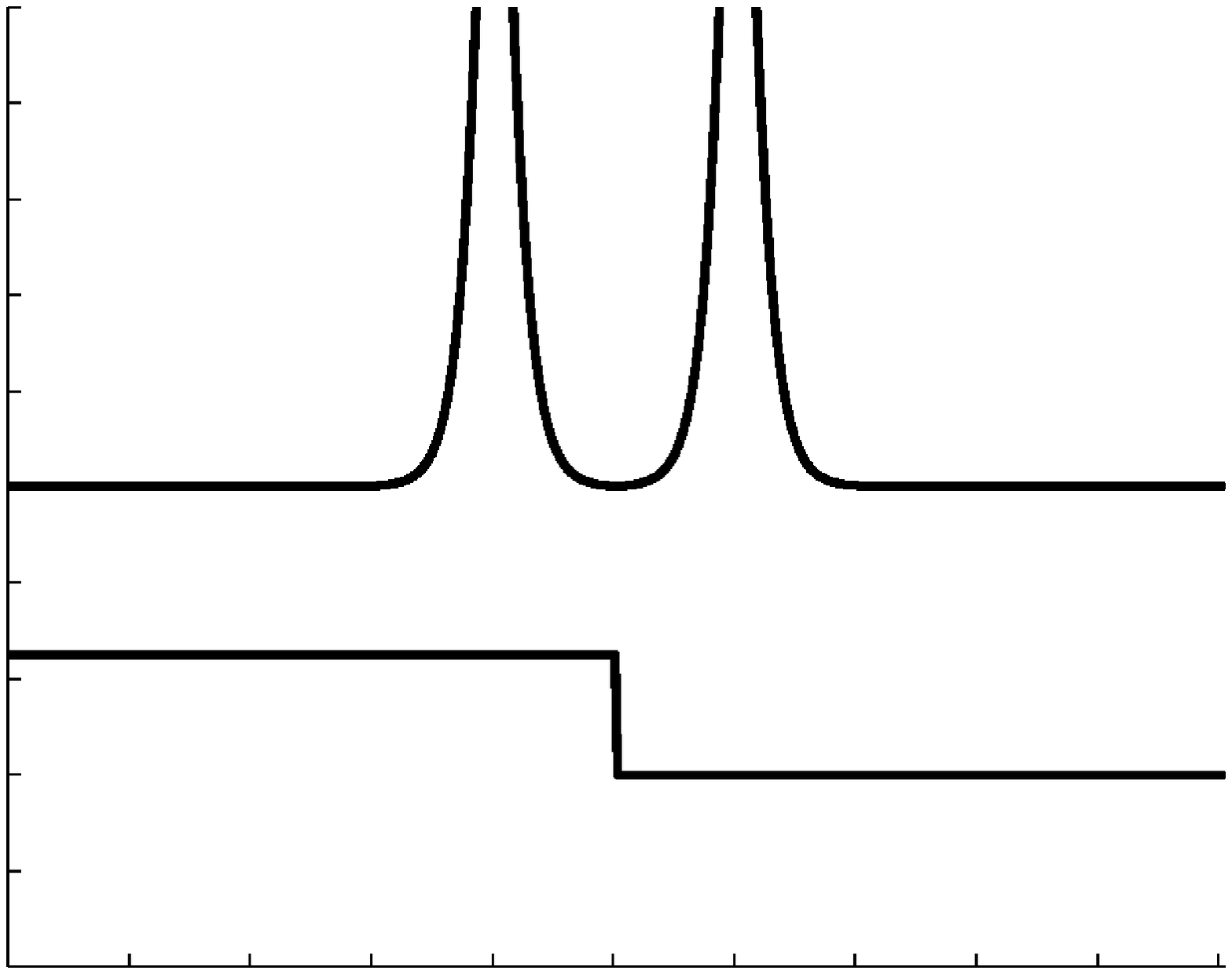}
\includegraphics[width=2.5cm,height=2.5cm]{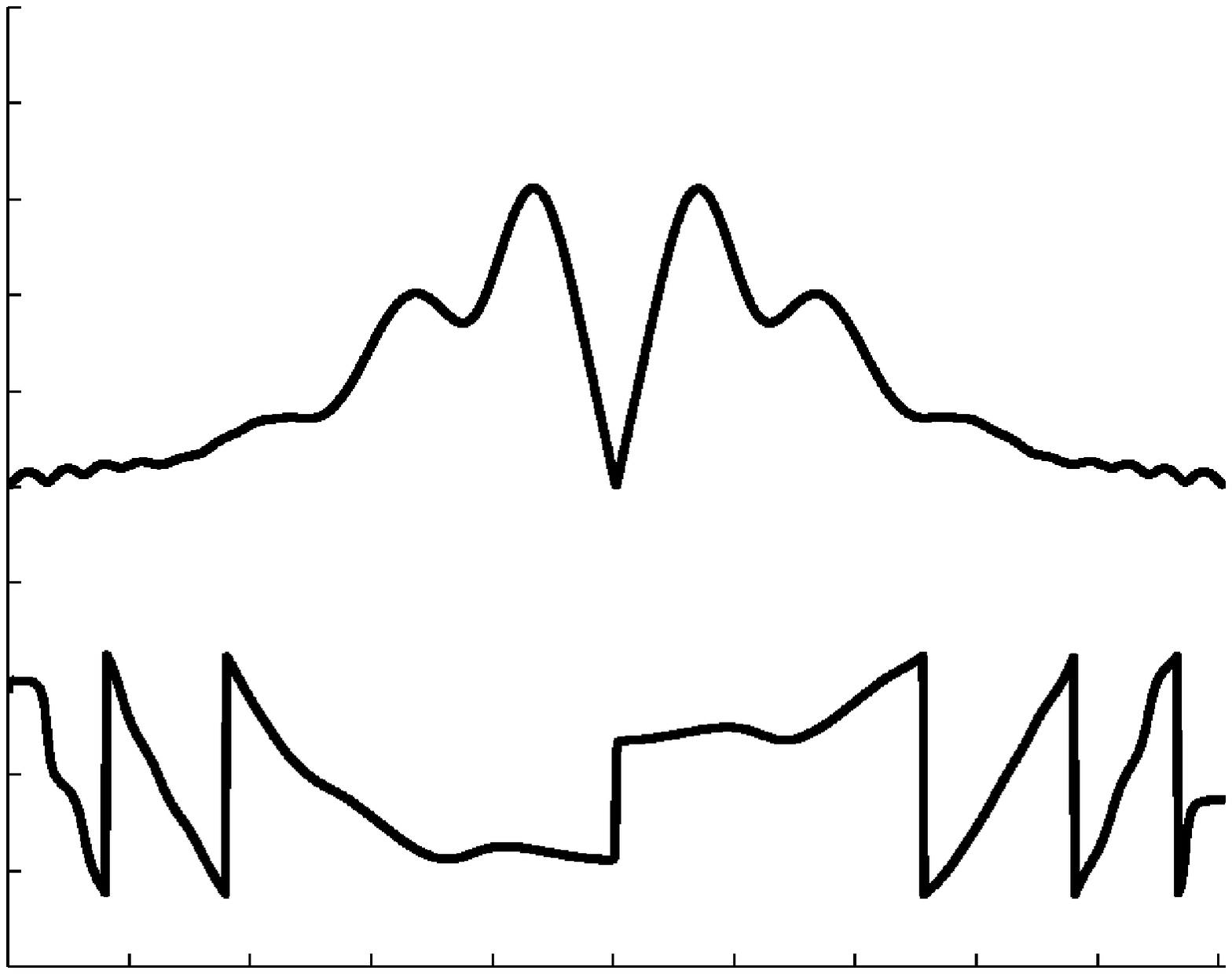}
\includegraphics[width=2.5cm,height=2.5cm]{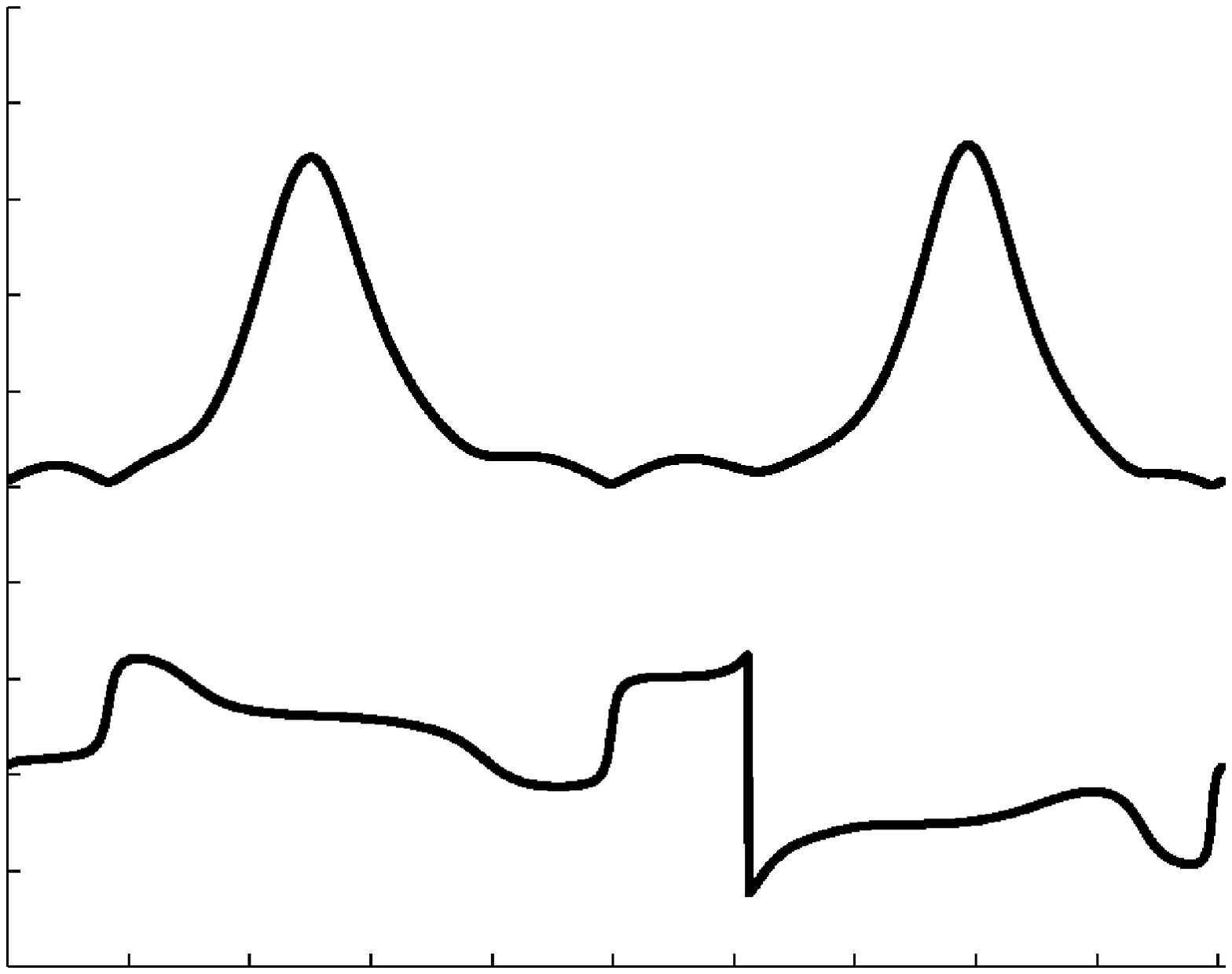}
\includegraphics[width=2.5cm,height=2.5cm]{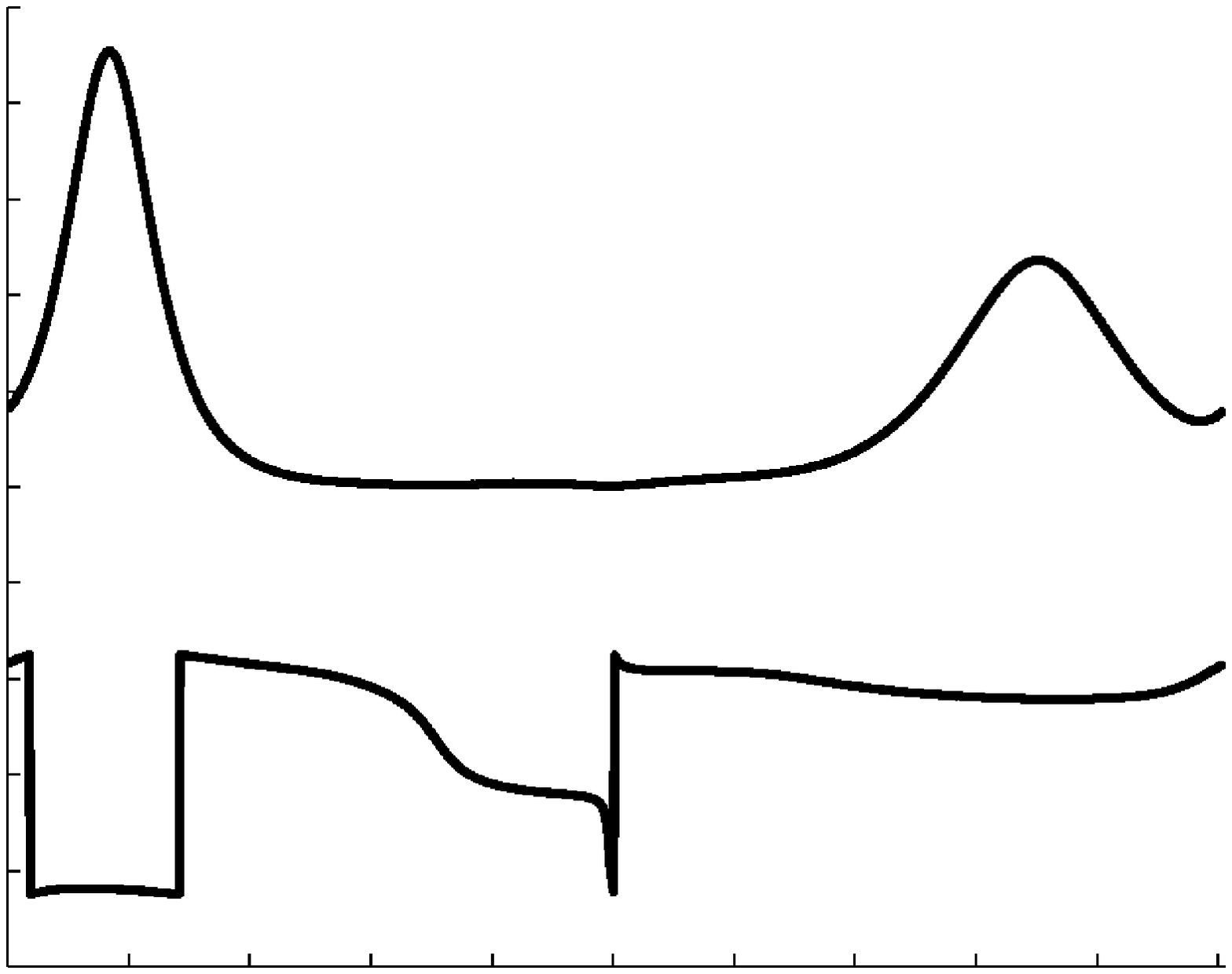}
\includegraphics[width=2.5cm,height=2.5cm]{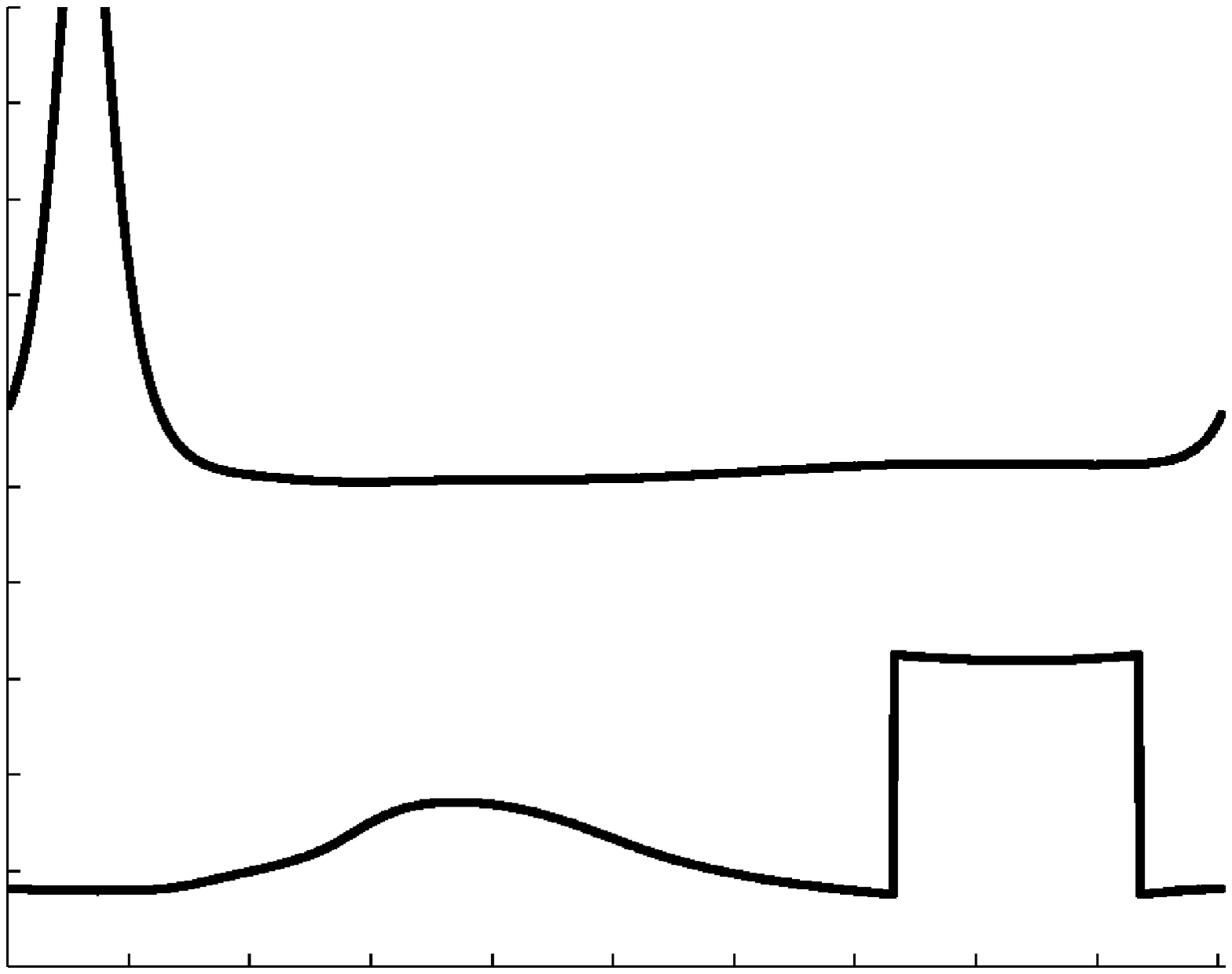}
\caption[]{Snapshots of $|\Psi(\theta, t)|$ (higher curve in
each panel) and of $\phi(\theta, t)$ (lower curve in each
panel) of the order parameter $\Psi =|\Psi(\theta,t)| e^{i
\phi(\theta,t)}$, for the antisymmetric initial condition, for
$t/t_0=0, 10, 100, 300$, and $400$, for a weak random potential
(shown in Fig.\,\ref{FIG14}), with an antisymmetric initial
condition, $\alpha = -1$. The axes are the same as in Fig.\,1.}
\label{FIG10}
\end{figure}

\begin{figure}[t]
\includegraphics[width=5.5cm,height=4.5cm]{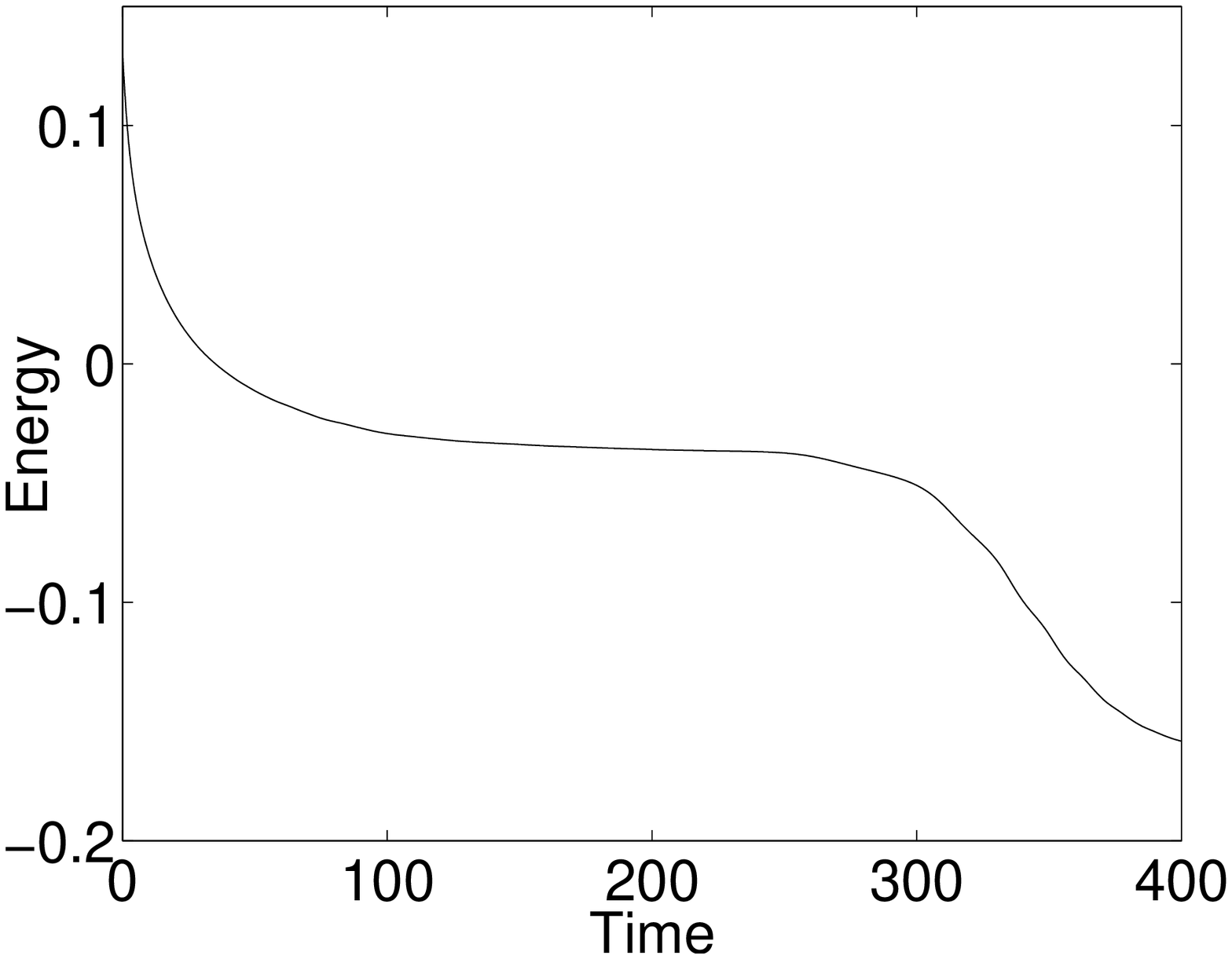}
\caption[]{The energy of the system in units of $E_0 =
\hbar^2/(2MR^2)$, versus time for $0 \le t/t_0 \le 400$
corresponding to Fig.\,\ref{FIG10}.}
\label{FIG11}
\end{figure}

\begin{figure}[t]
\includegraphics[width=2.5cm,height=2.5cm]{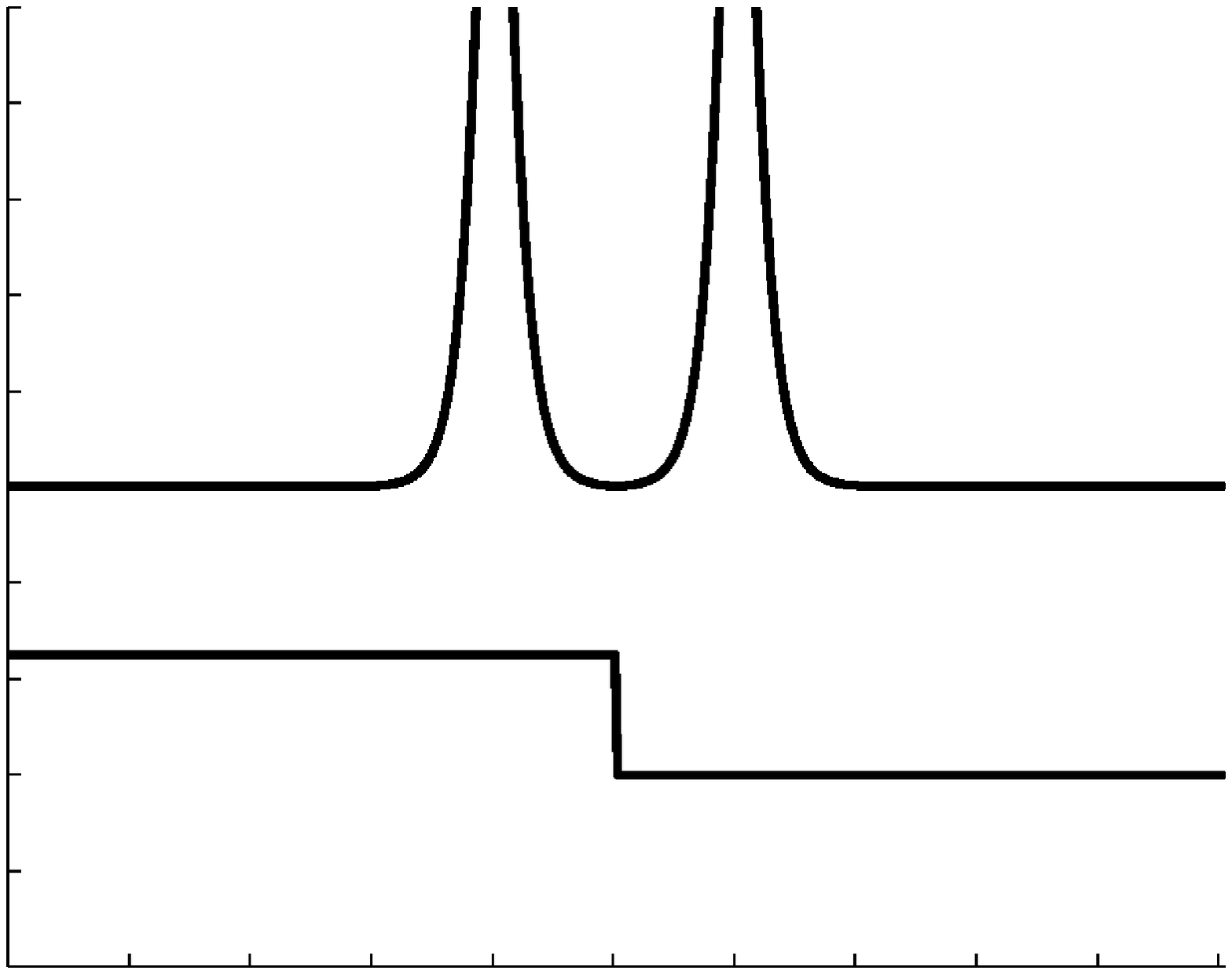}
\includegraphics[width=2.5cm,height=2.5cm]{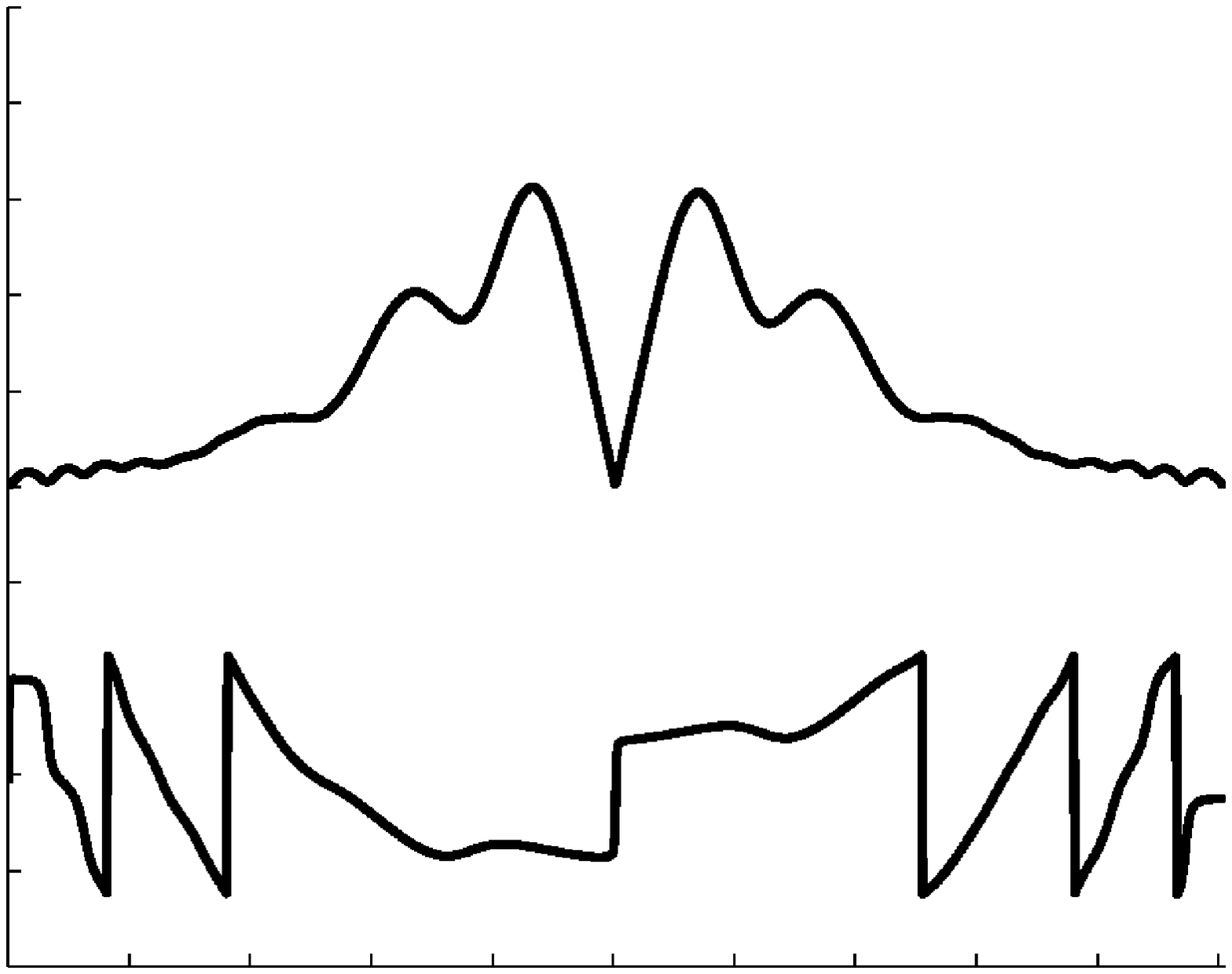}
\includegraphics[width=2.5cm,height=2.5cm]{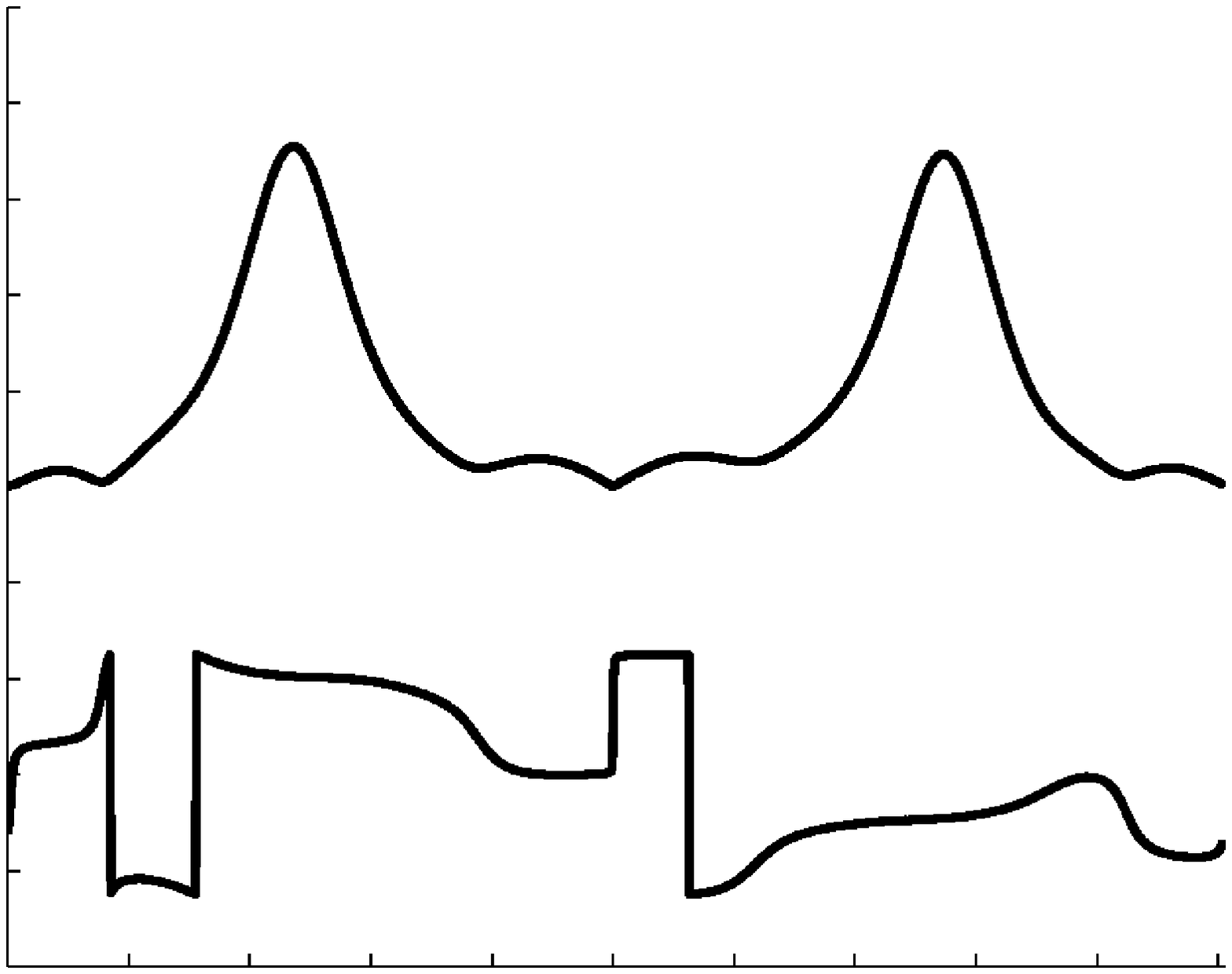}
\includegraphics[width=2.5cm,height=2.5cm]{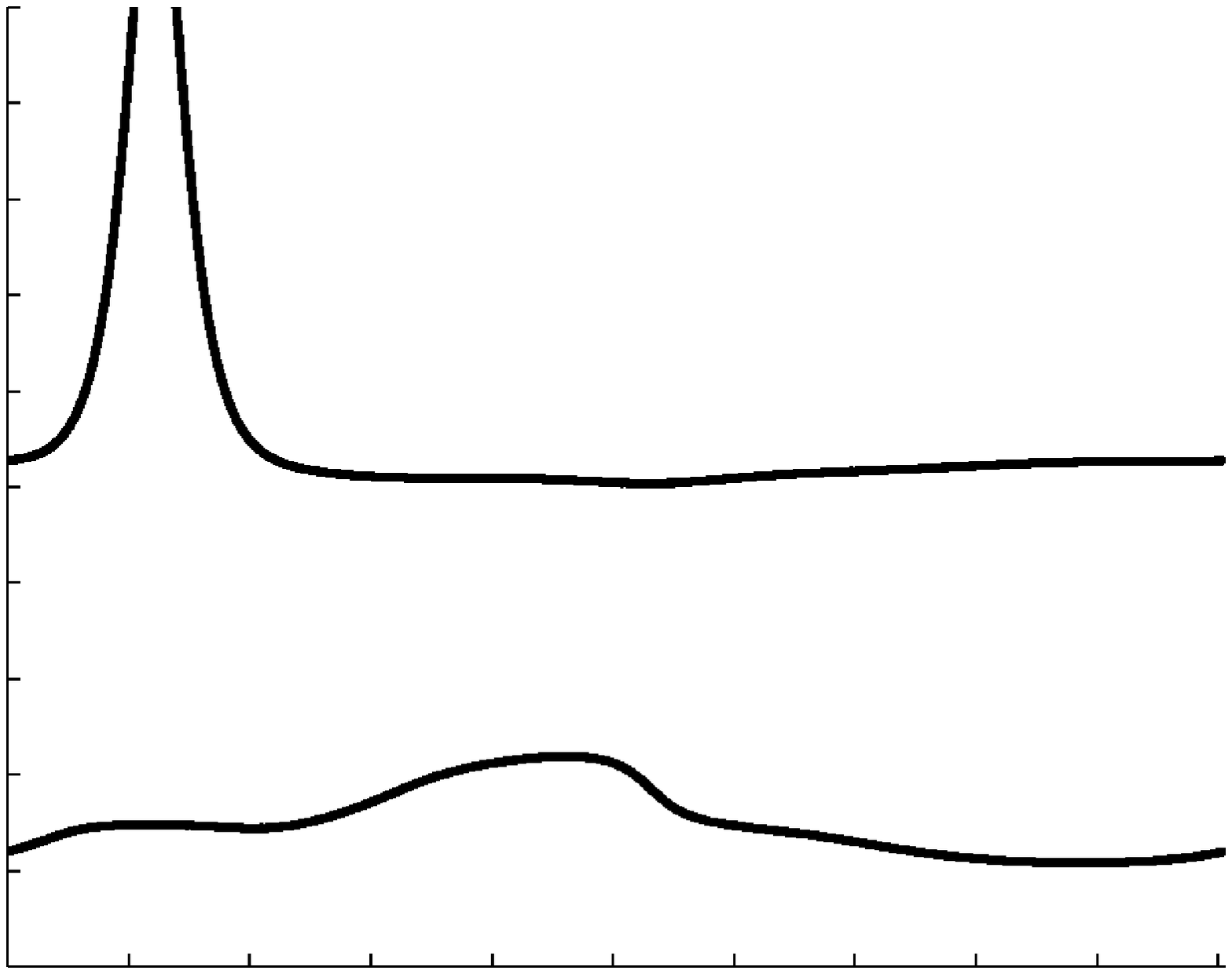}
\includegraphics[width=2.5cm,height=2.5cm]{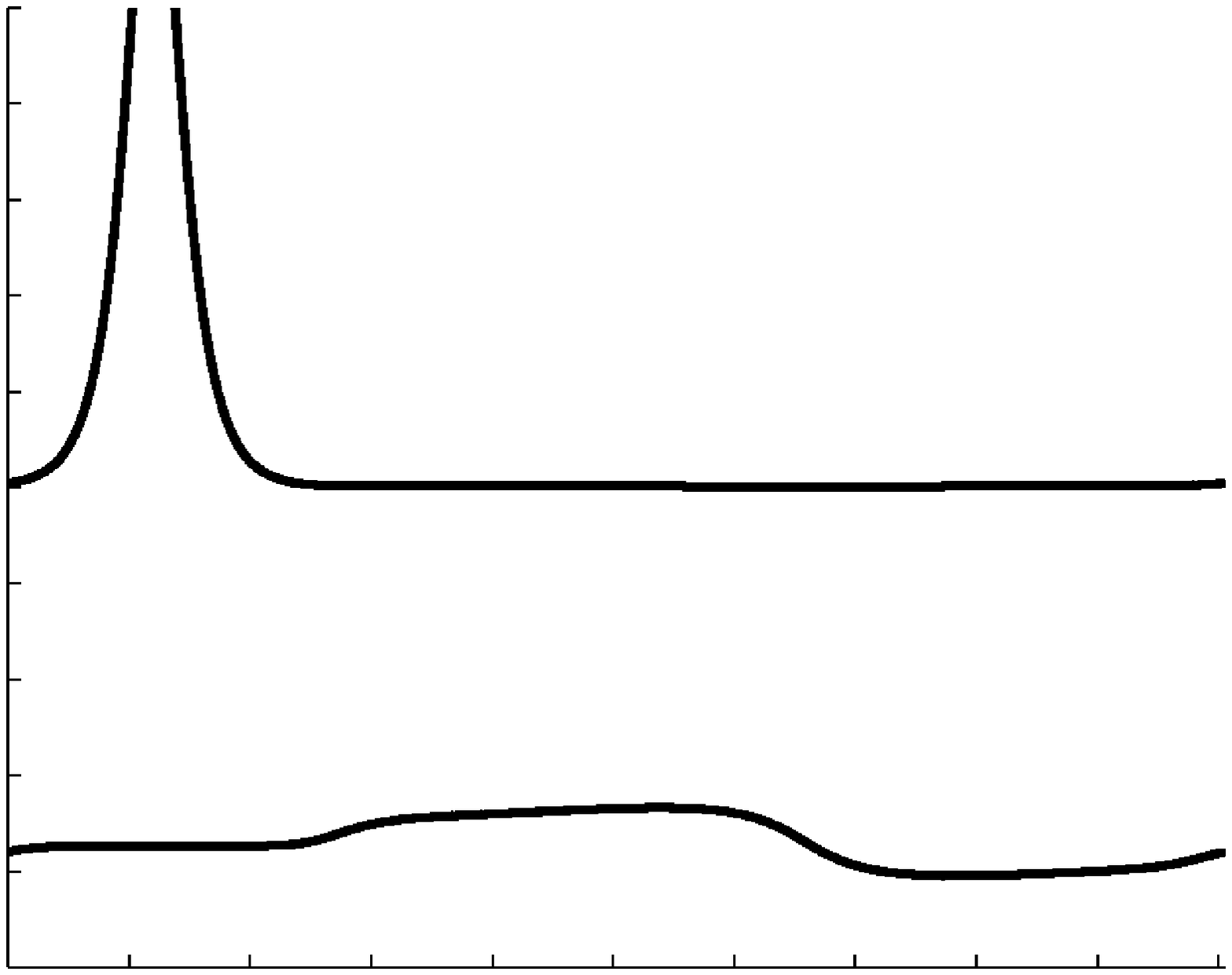}
\caption[]{Snapshots of $|\Psi(\theta, t)|$ (higher curve in
each panel) and of $\phi(\theta, t)$ (lower curve in each
panel) of the order parameter $\Psi =|\Psi(\theta,t)| e^{i
\phi(\theta,t)}$, for the almost antisymmetric initial
condition, $\alpha = -1.01$, for $t/t_0 = 0, 10, 100, 300$, and
$400$. The axes are the same as in Fig.\,1. In all the above
graphs there is no external potential, $V = 0$.}
\label{FIG12}
\end{figure}

\begin{figure}[t]
\includegraphics[width=5.5cm,height=4.5cm]{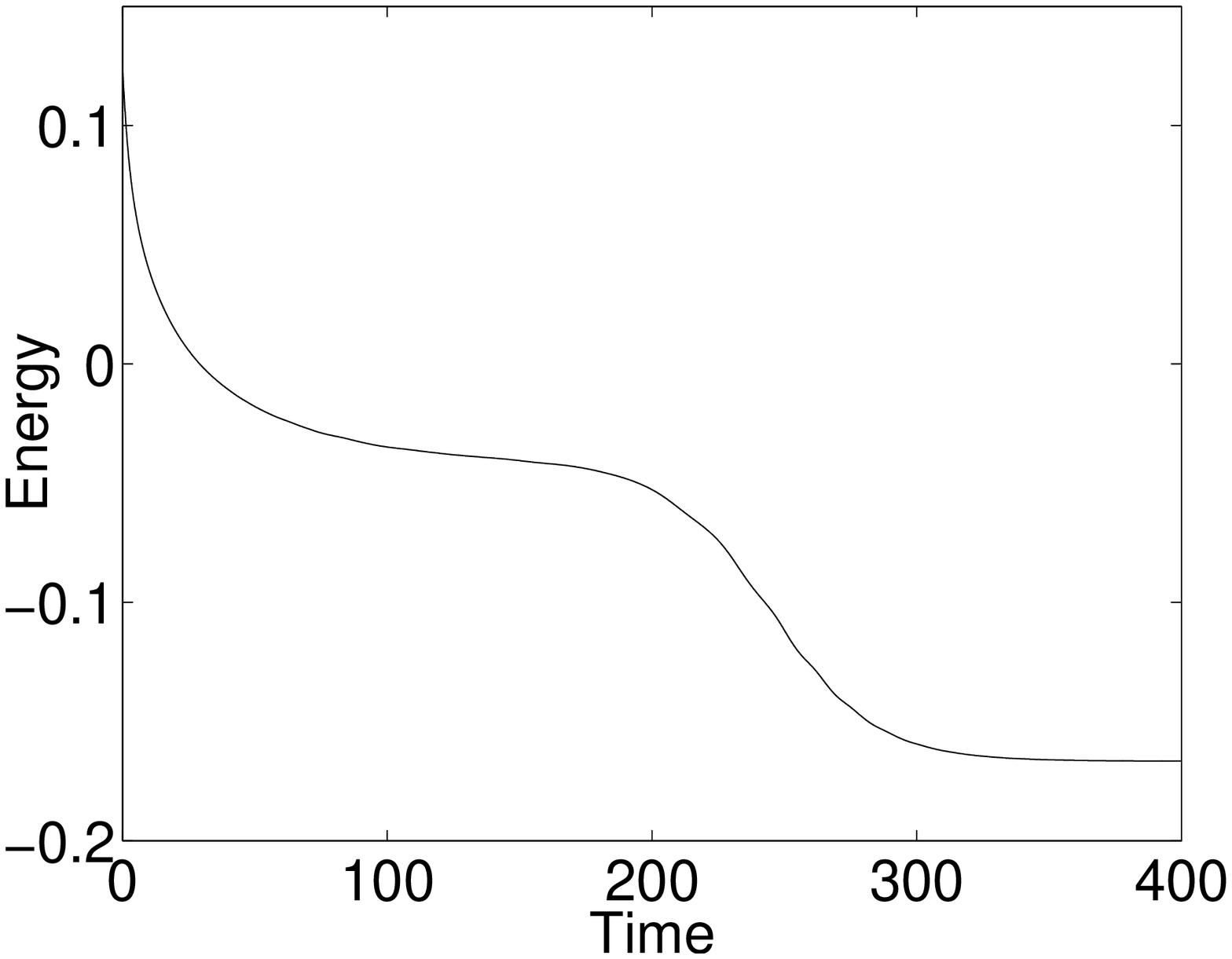}
\caption[]{The energy of the system in units of $E_0 =
\hbar^2/(2MR^2)$, versus time for $0 \le t/t_0 \le 400$
corresponding to Fig.\,\ref{FIG12}.}
\label{FIG13}
\end{figure}

Figures \ref{FIG2}, \ref{FIG4}, \ref{FIG6}, \ref{FIG8},
\ref{FIG10} and \ref{FIG12} show snapshots of $|\Psi(\theta,
t=0)|$ and of $\phi(\theta,t=0)$ for the order parameter
$\Psi(\theta, t) = |\Phi(\theta,t)| e^{i \phi(\theta, t)}$.
Figs.\,\ref{FIG3}, \ref{FIG5}, \ref{FIG7}, \ref{FIG9},
\ref{FIG11}, and \ref{FIG13} show the corresponding energy of
the gas as function of time.

\begin{figure}[t]
\includegraphics[width=8cm,height=5cm]{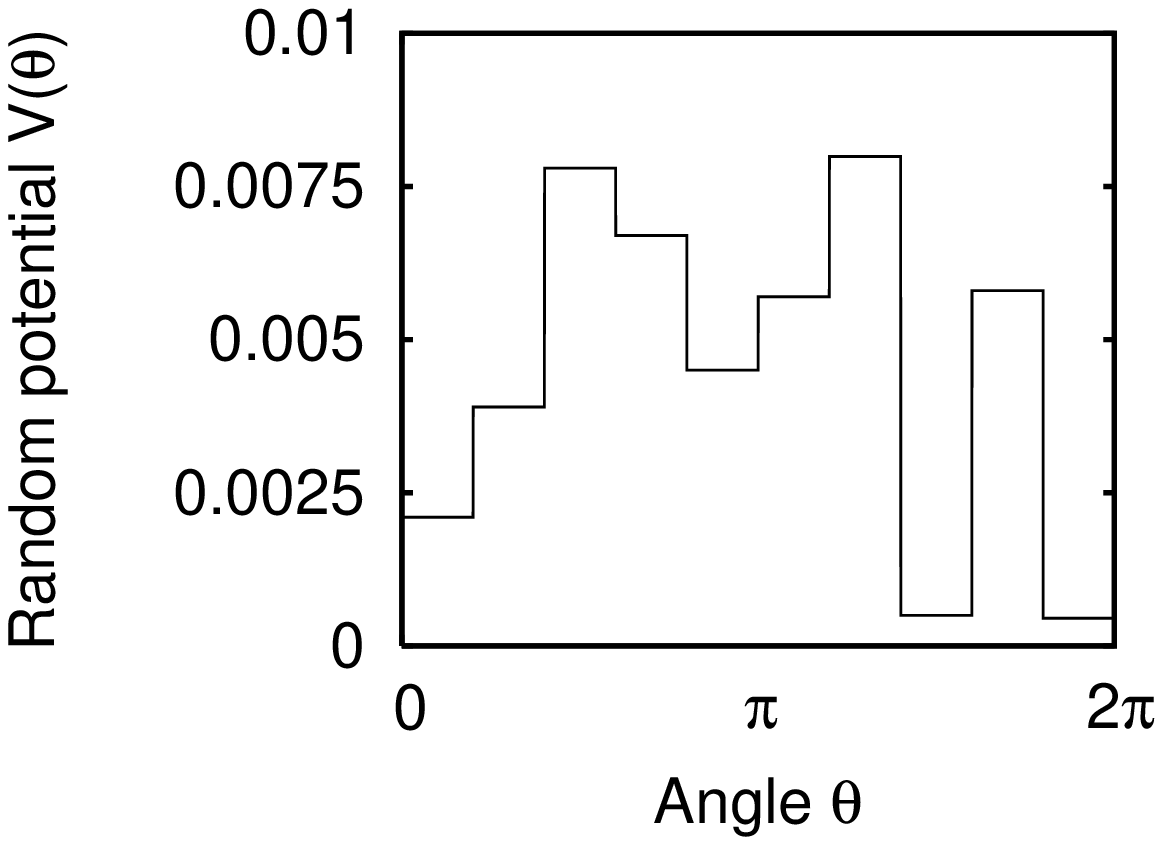}
\caption[]{The random potential in units of $E_0 =
\hbar^2/(2MR^2)$ as function of the angle $\theta$ that is used
in the specific calculations shown in Figs.\,\ref{FIG4} and
\ref{FIG10}.}
\label{FIG14}
\end{figure}

\section{Time evolution of the ``ideal" situation}

To understand the effects of a weak random potential and the
effect of slight asymmetries in the initial condition (to be
considered in Secs.\,IV and V), it is instructive to start with
the situation where there is no external potential, $V(\theta)
= 0$ and an initial configuration which is either perfectly
symmetric ($\alpha = 1$), or perfectly antisymmetric ($\alpha =
-1$), i.e.,
\begin{equation}
   \Psi(\theta, t=0) = \frac 1 {\sqrt{2}} [\psi(\theta - \theta_0)
   \pm \psi(\theta + \theta_0)].
\end{equation}
We also fix the value of the dissipative parameter equal to
$\gamma = 0.05$. Figures \ref{FIG2} and \ref{FIG8} show
snapshots of $|\Psi(\theta, t)|$ as well as the phase
$\phi(\theta, t)$ of the order parameter $\Psi(\theta, t)$ for
the symmetric and the antisymmetric case, respectively. The
snapshots shown in Fig.\,\ref{FIG2} correspond to $t/t_0= 0,
10, 50, 100$, and 150, and in Fig.\,\ref{FIG8} to $t/t_0=0, 10,
100, 300$, and $400$. Here $t_0 = E_0/\hbar = [\hbar/(2 M
R^2)]^{-1}$ is the unit of time. Figures \ref{FIG3} and
\ref{FIG9} show the energy of the system as function of time
for $0 \le t/t_0 \le 150$, and for $0 \le t/t_0 \le 400$,
respectively.

As seen in these graphs, the symmetric configuration
(Fig.\,\ref{FIG2}) merges quickly into one soliton and as time
increases, it eventually approaches the equilibrium solution.
On the other hand, the two blobs do not merge in the
antisymmetric case (Fig.\,\ref{FIG8}).  This is a direct
consequence of the fact that the initial configuration has a
node at $\theta = 0$. Because of the symmetry between $\theta$
and $-\theta$, $\Psi(\theta = 0, t)$ must be zero for all times
$t > 0$.  As a result, the two blobs never merge as a simple
consequence of parity conservation. The parity operator
commutes with the Hamiltonian, and parity is therefore a
conserved quantity. Only numerical errors can eventually lead
to a single-soliton profile (with lower energy).  That this
does not happen provides a check on the accuracy of our
numerics.

\section{Effect of the random potential on the time evolution}

Using Figs.\,\ref{FIG2} and \ref{FIG8} as ``reference plots",
we may now examine the effect of a weak, symmetry breaking
random potential $V(\theta)$. This potential is chosen to
consist of ten steps of equal widths with a height that is a
(uniformly distributed) random number and varies between 0 and
0.01. Figure \ref{FIG14} shows the specific random potential
chosen. In this case we start with perfectly
symmetric/antisymmetric configurations, $\alpha = \pm 1$.

The time evolution of the symmetric configuration shown in
Figs.\,\ref{FIG4} and \ref{FIG5} is almost identical with that
of Figs.\,\ref{FIG2} and \ref{FIG3}, i.e., the case considered
in the previous section with $V = 0$. The two blobs merge
rather rapidly.  The antisymmetric case shown in
Figs.\,\ref{FIG10} and \ref{FIG11} is of greater interest.
Here, after a relatively short time, the system passes through
a ``quasi-equilibrium" configuration, seen as the plateau in
the plot of energy versus time in Fig.\,\ref{FIG11}. During
this time interval, there are two localized blobs. However,
parity is no longer a conserved quantity in this case. There is
no symmetry in the system to preserve the node that was built
into the initial conditions.  As a result, the two blobs
eventually merge into one in contrast to the results of
Fig.\,\ref{FIG8}.  In other words, the apparent repulsion of
the two solitary waves is not present at sufficiently large
$t$.

\section{Effect of slight asymmetries in the initial configuration}

In another set of runs, we set the random potential to zero and
select slightly asymmetric initial configuration, with $\alpha
= \pm 1.01$. Our initial condition is thus not a parity
eigenstate and our calculations show that again the two
initially distinct blobs merge after a characteristic
timescale. The qualitative features of this calculation are the
same as in the case of a random potential described in the
previous section.

In the case of an almost symmetric initial configuration,
$\alpha = 1.01$ shown in Fig.\,\ref{FIG6}, the two separate
blobs merge rapidly, very much as in Figs.\,\ref{FIG2} and
\ref{FIG4}. On the other hand, the almost antisymmetric case,
$\alpha = -1.01$ shown in Figs.\,\ref{FIG12} and \ref{FIG13},
shows a plateau in the energy and a period of
``quasi-equilibrium" during which the two blobs have relatively
well-determined shape and location. Eventually, however, the
two blobs merge again into a single solitary wave, much as in
Fig.\,\ref{FIG10} but unlike Fig.\,\ref{FIG8}.

\section{Discussion and conclusions}

According to the results of our study, the observed repulsion
between bright solitary waves in the experiment of
Ref.\,\cite{sol3} implies that the conservation laws were not
substantially violated during the time interval investigated.
More precisely, it suggests that deviations from axial symmetry
in the trapping potential must have been small, the initial
configuration was very close to a (negative) parity eigenstate,
and that dissipation must have been weak.

It is instructive to estimate the timescale, $t_0$, for our
study. If one considers a value of $R$ equal to the
longitudinal size of an elongated trap, $R \sim 0.1$ mm, then
$t_0 \sim 10$ sec, which is a rather long timescale for these
experiments. Therefore, it seems likely that the characteristic
timescale over which the experiment of Ref.\,\cite{sol3} was
performed was significantly smaller than the timescale required
to see the separate blobs merge. Higher temperatures would
enhance the dissipation in the gas and would decrease the
characteristic time that is required for the blobs to merge. To
the extent that the deviations from axial symmetry in the
trapping potential and the antisymmetry in the initial
configuration considered here are representative of the actual
experimental situation, our results support the explanation
offered in Ref.\,\cite{Stoof}.  Direct experimental
determinations of these quantities and of the strength of
dissipation would thus be welcome. It would also be of interest
to investigate the long time stability (or instability) of the
configurations observed in Ref.\,\cite{sol3}.

The questions examined here may also have important
consequences on possible technological applications. For
example, propagation of such solitary waves in waveguides may
serve as signals that transfer energy or information.
Therefore, understanding and possibly controlling the way that
such waves interact with each other may be important. Recent
experimental progress in building quasi-one-dimensional and
toroidal traps should make such experiments easier to perform
and worth investigating.

\acknowledgements{We thank Magnus \"Ogren for useful
discussions.}


\begin{thebibliography}{99}

\bibitem{JKP} A. D. Jackson, G. M. Kavoulakis, and C. J.
Pethick, Phys. Rev. A {\bf 58}, 2417 (1998).

\bibitem{Sal} L. Salasnich, A. Parola, and L. Reatto, Phys. Rev. A
{\bf 65}, 043614 (2002).

\bibitem{sol1} S. Burger, K. Bongs, S. Dettmer, W. Ertmer,
K. Sengstock, A. Sanpera, G. V. Shlyapnikov, and M. Lewenstein,
Phys. Rev. Lett. {\bf 83}, 5198 (1999).

\bibitem{sol2} J. Denschlag, J. E. Simsarian, D. L. Feder,
Charles W. Clark, L. A. Collins, J. Cubizolles, L. Deng, E. W.
Hagley, K. Helmerson, W. P. Reinhardt, S. L. Rolston, B. I.
Schneider, and W. D. Phillips, Science {\bf 287}, 97 (2000).

\bibitem{sol3} K.~E.~Strecker, G.~B.~Partridge, A.~G.~Truscott,
and R.~G.~Hulet, Nature {\bf 417}, 150 (2002).

\bibitem{sol4} L.~Khaykovich, F.~Schreck, G.~Ferrari, T.~Bourdel,
J.~Cubizolles, L.~D.~Carr, Y.~Castin, and C.~Salomon, Science
{\bf 296}, 1290 (2002).

\bibitem{Stoof} U. Al Khawaja, H. T. C. Stoof, R. G. Hulet,
K. E. Strecker, and G. B. Partridge Phys. Rev. Lett. {\bf 89},
200404 (2002).

\bibitem{Carr} L. D. Carr and Y. Castin, Phys. Rev. A {\bf 66}, 063602
(2002).

\bibitem{Sal2} L. Salasnich, A. Parola, and L. Reatto, Phys. Rev. A,
{\bf 66}, 043603 (2002).

\bibitem{Gordon} J. P. Gordon, Optics Lett. {\bf 8}, 596 (1983).

\bibitem{CCR} L.~D.~Carr, C.~W.~Clark, and W.~P.~Reinhardt,
Phys. Rev. A {\bf 62}, 063611 (2000).

\bibitem{Kurn}  S. Gupta, K. W. Murch, K. L. Moore, T. P. Purdy,
and D. M. Stamper-Kurn, Phys. Rev. Lett. {\bf 95}, 143201
(2005).

\bibitem{phil} C. Ryu, M. F. Andersen, P. Clade, Vasant Natarajan,
K. Helmerson, W. D. Phillips, Phys. Rev. Lett. {\bf 99}, 260401 (2007).

\bibitem{Ueda} Rina Kanamoto, Hiroki Saito, and Masahito Ueda,
Phys. Rev. A {\bf 68}, 043619 (2003).

\bibitem{GMK} G. M. Kavoulakis, Phys. Rev. A {\bf 67}, 011601(R)
(2003).

\end{thebibliography}
\end{document}